\documentclass[10pt,preprint]{aastex61}

\usepackage{graphicx,amsmath,amssymb,amsbsy,bm,color,tabularx}
\usepackage{enumerate,isotope}
\usepackage{multirow,dcolumn}
\usepackage{hhline}
\usepackage{bbm}
\allowdisplaybreaks[1]

\newcommand{\fmiq}{\, \text{fm}^{-3}}

\newcommand{\mev}{\, \text{MeV}}

\newcommand{\nxlo}[1]{%
   \ifnum0=#1\relax%
      \text{LO}%
   \else%
   \ifnum1=#1\relax%
      \text{NLO}%
   \else%
      \text{N}\ensuremath{^{#1}}\text{LO}%
   \fi
   \fi
}

\newcommand{\beq}{\begin{equation}}
\newcommand{\eeq}{\end{equation}}

\definecolor{dartmouth}{rgb}{0.05,0.5,0.06}

\allowdisplaybreaks

\begin{document}

\title{Constraining the speed of sound inside neutron stars \\with chiral effective field theory interactions and observations}
\author{I.\ Tews}
\email{itews@uw.edu}
\affiliation{
Institute for Nuclear Theory, University of Washington, Seattle, WA 98195-1550, USA}
\affiliation{JINA-CEE, Michigan State University, East Lansing, MI, 48823, USA}

\author{J.\ Carlson}
\email{carlson@lanl.gov}
\affiliation{Theoretical Division, Los Alamos National Laboratory, Los Alamos, NM 87545, USA}
\author{S.\ Gandolfi}
\email{stefano@lanl.gov}
\affiliation{Theoretical Division, Los Alamos National Laboratory, Los Alamos, NM 87545, USA}
\author{S.\ Reddy}
\email{sareddy@uw.edu}
\affiliation{
Institute for Nuclear Theory, University of Washington, Seattle, WA 98195-1550, USA}
\affiliation{JINA-CEE, Michigan State University, East Lansing, MI, 48823, USA}

\begin{abstract}
The dense matter equation of state (EOS) determines neutron star (NS) structure 
but can be calculated reliably only up to one to two times the nuclear saturation density, using 
accurate many-body methods that employ nuclear interactions from chiral effective
field theory constrained by scattering data. In this work, we use physically 
motivated ansatzes for the speed of sound $c_S$ at high density to extend 
microscopic calculations of neutron-rich matter to the highest densities encountered 
in stable NS cores. We show how existing and expected astrophysical 
constraints on NS masses and radii from X-ray observations can constrain the 
speed of sound in the NS core. We confirm earlier expectations that $c_S$ is likely 
to violate the conformal limit of $c_S^2\leq c^2/3 $, possibly reaching 
values closer to the speed of light $c$ at a few times the nuclear saturation density, 
independent of the nuclear Hamiltonian. If QCD obeys the 
conformal limit, we conclude that the rapid increase of $c_S$ required to 
accommodate a $2 $ M$_\odot$ NS suggests a form of strongly interacting matter 
where a description in terms of nucleons will be unwieldy, even between one and two times the
nuclear saturation density. For typical NSs with masses in the range 
$1.2-1.4~$ M$_\odot$, we find radii between $10$ and $14$ km, and the smallest possible 
radius of a $1.4$ M$_{\odot}$ NS consistent with constraints from nuclear physics 
and observations is $8.4$ km. We 
also discuss how future observations could constrain the EOS and guide theoretical 
developments in nuclear physics.

\end{abstract}

\section{Introduction}

The neutron-matter equation of state (EOS) at $T=0$ is a crucial ingredient 
to describe the structure of neutron stars (NS), as expressed by the 
mass-radius relation.  At the high densities in the NS core, the proton 
fractions are small, typically $5-10$\%, and NS matter can 
be effectively described as pure neutron matter (PNM). Although protons are
included when constructing the EOS of NSs, their contribution is small 
(compared to other uncertainties, which we shall discuss in some detail) and 
always acts to reduce the pressure.  
Up to $\sim1-2$ times the nuclear saturation
density, $n_0=0.16 \fmiq \simeq 2.7\cdot 10^{14} \rm{g}\, \rm{cm}^{-3}$, the 
EOS of PNM has been computed by using different many-body 
methods with realistic nucleon-nucleon interactions; see, e.g.,~\cite{Hebeler:2009iv, Gandolfi:2011xu,Tews:2012fj, Hagen:2013yba, Gandolfi:2015jma, Sammarruca:2014zia, Lynn:2015jua, Drischler:2016djf} and \cite{Holt:2016pjb}. 
In the following, we shall collectively refer to them as ab initio calculations. 
At densities above $2 n_0$, however, the PNM EOS is not constrained.

From the EOS, $p=p(\epsilon)$, we can obtain the speed of sound $c_S$
from
\begin{align}
c_S^2=\frac{\partial p(\epsilon)}{\partial \epsilon}\,,
\end{align}
with the pressure $p$ and the energy density $\epsilon$, where the latter
also includes the rest-mass contribution. Stability and causality constrain 
$0\leq c_S^2 \leq 1$, where we have set the speed of light $c=1$. Current ab 
initio calculations of 
neutron matter constrain the speed of sound up to $\approx 1-2 n_0$, but, again,
at higher densities uncertainties grow rapidly and the speed of sound 
remains unconstrained. 

In this work, we show that useful constraints on the speed of sound at  higher 
densities can be deduced from the observation of two-solar-mass NSs~\citep{Demorest2010, Antoniadis2013} by using ab initio calculations of the 
EOS of PNM up to $1-2 n_0$. Our work is similar in some respects to earlier work 
by \cite{Bedaque:2014sqa},
where $c_S$ at high density was constrained using a parameterized EOS and astrophysical observations. An important distinction is that we use ab initio methods and nuclear interactions from chiral effective field theory (EFT) which allows us to study the influence of uncertainties due to poorly constrained short-distance behavior of two- and three-nucleon interactions.   

At very high densities, far exceeding the densities in the NS core, additional information on the speed of sound can be obtained from perturbative QCD.
These calculations of cold ultra-dense quark matter, which are reliable at asymptotic density, show that corrections due to interactions between quarks decrease $c_S$ and that $c_S^2 <1/3$~\citep{Kurkela:2009gj}. 
They also show that  $c_S^2 \rightarrow 1/3$ with increasing density from below.   
In general, in conformal theories, where the trace of the energy momentum tensor $\epsilon - 3P$ vanishes, $c_S^2 =1/3$ is independent of density, temperature, or interactions.
Lattice QCD calculations at finite temperature and zero chemical potential show that $c_S^2 <1/3$, and that the introduction of a small baryon chemical potential does not alter this result \citep{Borsanyi:2012cr}. The speed of sound has also been calculated in a large class of theories for which the ADS/CFT correspondence holds, and calculations are possible in the strong coupling limit. It has been conjectured that $c_S^2 <1/3$, even when the trace of the energy momentum tensor is nonzero \citep{Cherman:2009tw}, although recently explicit counterexamples have been presented \citep{Hoyos:2016cob, Ecker:2017fyh}. We refer the reader to \cite{Bedaque:2014sqa} for a more detailed discussion of this conjecture and known exceptions (see also \cite{Hoyos:2016cob} and \cite{Ecker:2017fyh}). 

\begin{figure}[t]
\centering
\includegraphics[trim= 0 0 0 0cm, clip=,width=0.7\textwidth]{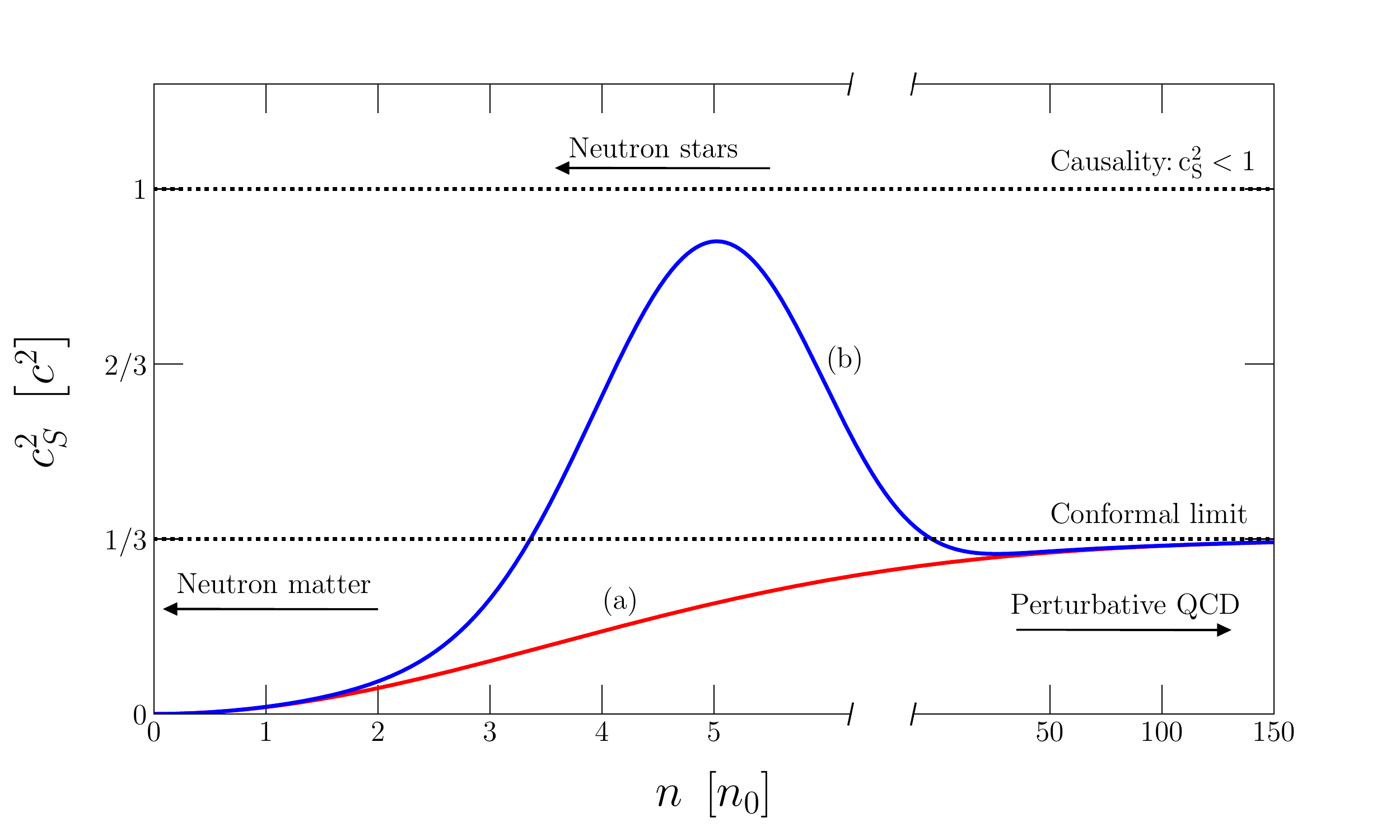} 
\caption{Two possible scenarios for the evolution of the speed of sound in dense matter. \label{fig:cS2_illustrate}}
\end{figure}

For QCD at finite baryon density, we are unaware of compelling reasons to expect that $c_S^2 <1/3$, and based on the preceding arguments, we will consider two minimal scenarios, which are illustrated in Fig.~\ref{fig:cS2_illustrate}. The scenario labeled 
(a) corresponds to the case when we assume that QCD obeys the conformal limit $c_S^2 <1/3$ at all densities, and scenario (b) corresponds to QCD violating this conformal bound. The behavior of $c_S$ at low and high density is constrained by theory, and we shall show that NS observations, when combined with improved ab initio calculations of PNM, can distinguish between these two scenarios, and provide useful insights about matter at densities realized inside NSs. 

This paper is structured as follows. In Section~\ref{sec:csnucl}, we present 
constraints on the speed of sound from nuclear physics. In
Section~\ref{sec:EOSextension}, we extend the speed of sound to higher 
densities. In Section~\ref{sec:csconformal}, we study the EOS under the assumption 
that the conformal limit is obeyed and the speed of sound is bounded by 
$1/\sqrt{3}$. For this case, we find that $c_S$ needs to increase very rapidly above 
 $1-2 n_0$ to stabilize a $2$ M$_\odot$ NS. Such a rapid increase likely signals the appearance of a new form of strongly coupled matter where the nucleon is no 
longer a useful degree of freedom. In Section~\ref{sec:csnonconf}, we release this assumption but still find that models in which $c_S$ increases rapidly, reaching values 
close to $c$, are favored. We study correlations in our parameterization in Section~\ref{sec:correlations}. 
In Section~\ref{sec:minradius}, we derive the smallest possible radius for NSs consistent with nuclear physics and observations. We then investigate 
the impact of possible additional observations in Section~\ref{sec:obs}. 
Finally, we summarize our main findings in Section~\ref{sec:summary}.

\section{EOS and speed of sound from nuclear physics}\label{sec:csnucl}

\begin{figure}[h]
\centering
\includegraphics[trim= 0 0 0 0, clip=,width=0.6\textwidth]{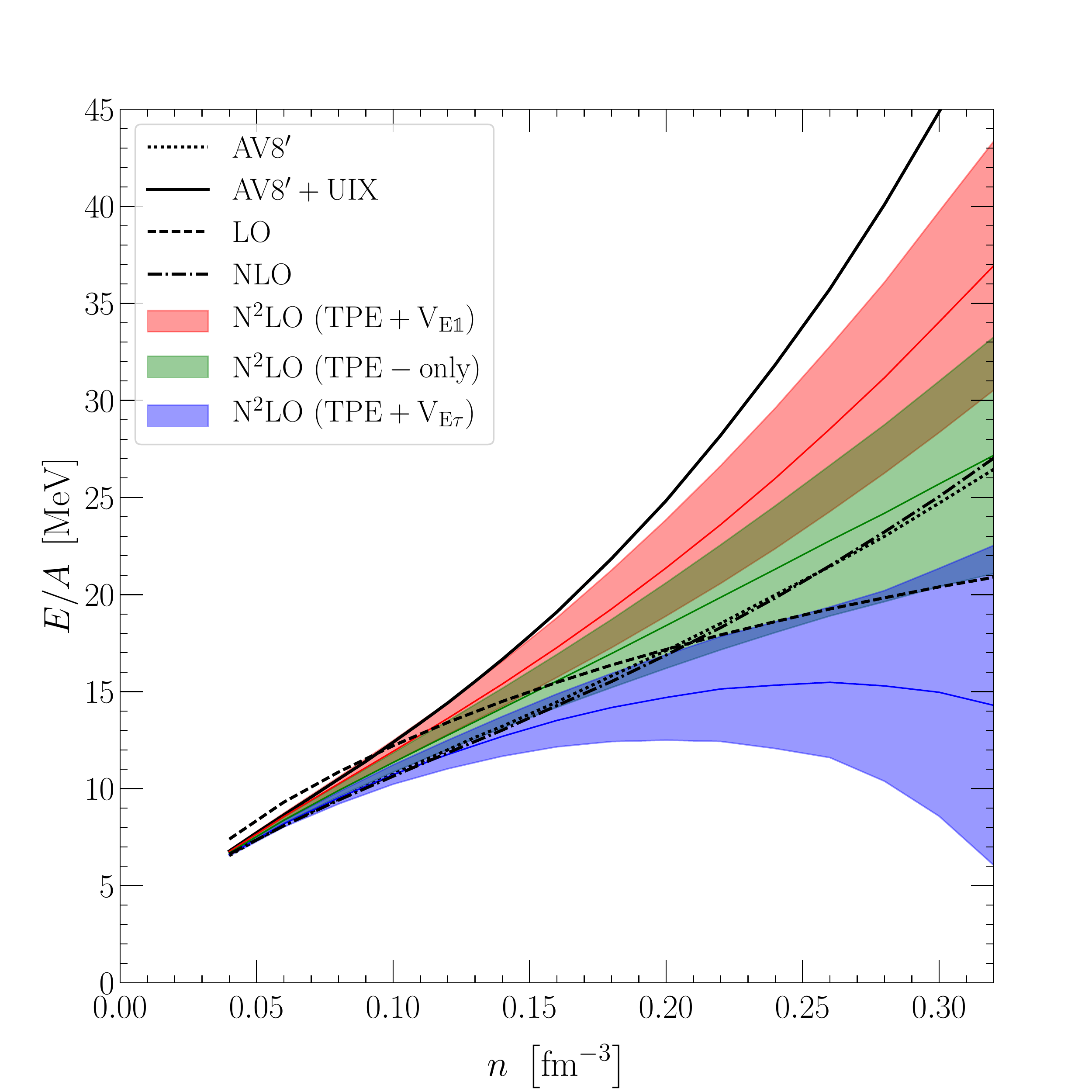} 
\caption{Neutron-matter EOSs used in this work. We show the AFDMC results 
for local chiral Hamiltonians with three different $3N$ short-range operators:
TPE-only (green middle band), 
TPE+$V_{E,\mathbbm{1}}$ (red upper band), and TPE+$V_{E,\tau}$ (blue lower band),
see~\cite{Lynn:2015jua} for details. For comparison, we also show results
for the phenomenological AV8'+UIX interactions (black line), for AV8' (dotted line), as well as LO (dashed
line) and NLO (dashed-dotted line) results for the local chiral interactions 
of~\cite{Gezerlis:2014zia} with $R_0=1.0$ fm.
\label{fig:nmeos}}
\end{figure}

\subsection{The EOS of neutron matter}

In this work, we use auxiliary-field diffusion Monte Carlo (AFDMC) to find the many-body ground state for a given nonrelativistic nuclear 
Hamiltonian~\citep{Carlson:2014vla}. In general, the nuclear Hamiltonian
contains two-body ($NN$), three-body ($3N$), and higher many-body ($AN$) 
forces,
\begin{align}
\mathcal{H}= \mathcal{T}+\mathcal{V}_{\text{NN}}+\mathcal{V}_{\text{3N}}
+\mathcal{V}_{\text{AN}}\,, 
\end{align}
which can be obtained from chiral effective field theory (EFT) at low-density (see, 
for instance,~\cite{Epelbaum:2008ga} and \cite{Machleidt:2011zz}).
Chiral EFT is a systematic framework for low-energy hadronic interactions, 
that naturally includes both two-body and many-body forces and allows for 
systematic uncertainty estimates. It has been successfully used to calculate 
nuclei and nuclear matter, see, for instance, \cite{Hebeler:2015hla} and 
references therein. 

In this paper, we extend the AFDMC calculations of PNM of 
\cite{Lynn:2015jua} with recently developed local chiral N$^2$LO interactions, 
including two- and three-body forces of~\cite{Gezerlis:2013ipa}, \cite{Gezerlis:2014zia} and \cite{Tews:2015ufa}, to higher densities. 
We find that, despite the rapid increase of the error estimates, EFT-based 
interactions remain useful up to $n=0.32 \fmiq$ and our results for the energy per 
particle in neutron matter are shown in Figure~\ref{fig:nmeos}.  
We plot the results for local chiral interactions at LO, NLO, and N$^2$LO with three different $3N$ 
interactions defined in~\cite{Lynn:2015jua}: 
$3N$ interactions with only the two-pion exchange (TPE-only), and $3N$ 
interactions containing the TPE  plus shorter-range contact terms with two different 
spin-isospin operators (TPE+$V_{E,\mathbbm{1}}$ and TPE+$V_{E,\tau}$), 
see~\cite{Lynn:2015jua} for details. The  uncertainty bands for the 
individual N$^2$LO interactions are obtained as suggested by~\cite{Epelbaum:2014efa}, 
i.e., the uncertainty $\Delta X^{\nxlo{2}}$ at order N$^2$LO is given by
\begin{align}
\Delta X^{\nxlo{2}}=\max
\left(\vphantom{X^{\nxlo{2}}}Q^{4} \left|X^{\nxlo{0}}-X^{\rm free}\right|,Q^2 \left|X^{\nxlo{1}}-X^{\nxlo{0}}\right|,
Q\left|X^{\nxlo{2}}-X^{\nxlo{1}}\right|
\right)\,, \label{eq:uncertainty}
\end{align}
and similar for lower orders. The dimensionless expansion parameter $Q$ is given by 
$Q=\max(p/\Lambda_B,m_{\pi}/\Lambda_B)$ with $p$ being a typical momentum 
scale of the system, $m_{\pi}$ being the pion mass, and $\Lambda_B\approx 500$ MeV being the breakdown scale of chiral EFT. For PNM in Figure~\ref{fig:nmeos}, we use the scale 
$p=\sqrt{3/5} k_F$; see~\cite{Lynn:2017fxg}. The results indicate that the EOS of 
PNM is well constrained up to saturation 
density, but the associated error grows rapidly with density. Nonetheless, the 
order-by-order convergence is still consistent with EFT expectations in the 
range $n_0-2n_0$.  In addition, although these interactions are only fit to low-energy 
scattering data up to laboratory energies of $150$ MeV, they also describe phase 
shifts at much higher energies within uncertainties; see, e.g., 
\cite{Epelbaum:2014efa}. For comparison, we also show results obtained using the 
phenomenological Argonne v8' $NN$ interaction (AV8') and the Urbana IX $3N$ 
interactions (UIX) in Figure~\ref{fig:nmeos}.  

We present the energy per particle and the pressure at
$n_0$ and $2n_0$ at LO, NLO, and N$^2$LO in Table~\ref{tab:convergence} to 
provide supporting arguments for the convergence of the chiral expansion in this
density range. Uncertainty estimates in Table~\ref{tab:convergence} are provided 
by assuming a typical momentum scale $p=\sqrt{3/5} k_F$ and, in addition, a
more conservative choice $p=k_F$.  
At saturation density, we find a systematic order-by-order convergence for both
energy and pressure: the results at different orders overlap within their
 uncertainties, which decrease order-by-order.

\begin{table}
\caption{\label{tab:convergence}
Energy per particle and pressure in PNM at $n_0$ and $2n_0$ for local chiral 
Hamiltonians at LO, NLO, and N$^2$LO with three different short-range 
operators. The uncertainties are obtained using Equation~\eqref{eq:uncertainty} 
with two different expansion parameters $p/\Lambda_B$ with $p=\sqrt{3/5} k_F$ 
and $p=k_F$ in parenthesis.
We also give the values for the free Fermi gas and the phenomenological 
AV8'+UIX interaction.}
\scriptsize{
\begin{ruledtabular}
\begin{tabular}{cccc|ccccc}
& & free & pheno. & LO & NLO & N$^2$LO (TPE-only) & N$^2$LO (+
$V_{E,\mathbbm{1}}$) &  N$^2$LO (+$V_{E,\tau}$)\\
\hline
E/A & $n_0$ & 35.1 & 19.1 & $15.5 \pm 5.2\, (8.6)$ & $14.3 \pm 2.7\, (5.7)$ & $15.6 \pm 1.4\, (3.8)$ & $17.3 \pm 1.5\, (3.8)$ & $13.5 \pm 1.4\, (3.8)$ \\
& $2n_0$  &  55.7 & 49.9 & $20.9 \pm 14.6\, (24.3)$ & $27.0 \pm 9.4\, (20.3)$ & $27.2 \pm 6.1\, (16.9)$ & $36.9 \pm 6.4\, (16.9)$ & $14.3 \pm 8.2\, (16.9)$\\
P &$n_0$  & 3.7 & 3.3 & $1.3 \pm 0.7\, (1.1)$ & $1.6 \pm 0.4\, (0.8)$ & $1.8 \pm 0.2\, (0.5)$ & $2.4 \pm 0.4\, (0.6)$ & $1.1 \pm 0.3\, (0.5)$ \\
&$2n_0$  & 11.9 & 25.8 &$3.1 \pm 3.7\, (6.1)$ & $9.8 \pm 4.4\, (5.6)$ & $7.8 \pm 2.8\, (4.7)$ & $15.1 \pm 3.4\, (4.7)$ & $-2.6 \pm 8.1\, (10.4)$
\end{tabular}
\end{ruledtabular}}
\end{table}

From Figure~\ref{fig:nmeos} and Table~\ref{tab:convergence}, we see that the 
shorter-range $3N$ interactions significantly influence the convergence, but are still 
within conservative uncertainty estimates. These shorter-range $3N$ forces appear 
at N$^2$LO in the chiral power counting and contribute to systems containing 
triples of both neutrons and protons with $S=1/2$ and $T=1/2$, where $S$ and 
$T$ are the total spin and isospin, respectively. They are typically fit to few-body 
observables, such as the $\isotope[4]{He}$ binding
energy and neutron-alpha scattering (see \cite{Lynn:2017fxg} for a detailed 
discussion). 
In PNM, where all triples have $T=3/2$, these $3N$ forces usually vanish and it can 
be shown that at N$^2$LO only the long-range $3N$ TPE interaction contributes (see, for instance, \cite{Hebeler:2009iv}), while shorter-range contributions to PNM 
only appear at higher-order in the chiral expansion. 
In our case, however, contributions to PNM from short-range N$^2$LO 
$3N$  interactions arise as artifacts from using local 
regulators. Local forces in coordinate space are essential for their incorporation in 
AFDMC but introduce regulator artifacts; see~\cite{Huth:2017wzw} for a detailed 
discussion in the $NN$ sector. In the case of three-nucleon forces, these regulator 
artifacts appear in form of shorter-range $3N$ interactions mixed into 
triples with $S=3/2$ or $T=3/2$, e.g., triples containing three neutrons. We 
include these additional contributions, which we denote as N$^2$LO 
TPE+$V_{E,\mathbbm{1}}$ and N$^2$LO TPE+$V_{E,\tau}$, to provide even 
more conservative uncertainty estimates.   
 
It is interesting to analyze the predictions for the energy per particle and the 
pressure of PNM at $2n_0$. From Table~\ref{tab:convergence} and comparing 
the LO, NLO, and N$^2$LO TPE-only predictions, the pattern of convergence for 
the energy per particle is quite reasonable, with clear signs of an 
order-by-order improvement. For the pressure, the NLO contribution is larger 
than expected from naive power counting arguments. A plausible resolution of 
this discrepancy could be important contributions due to the $\Delta$ isobar, which 
are only included at N$^2$LO in $\Delta$-less chiral EFT. We note, however, that 
the more 
conservative error estimates in brackets assuage the tension somewhat. The 
improvement in predictions for the pressure at $2n_0$ from  NLO to N$^2$LO 
(TPE-only), on the other hand, is consistent with EFT expectations.  The situation 
for the TPE+$V_{E,\mathbbm{1}}$  and TPE+$V_{E,\tau}$ interactions is less 
satisfactory. For the TPE+$V_{E,\mathbbm{1}}$ interaction, the predicted pressure 
at N$^2$LO has some overlap with the error band at NLO, while 
for the TPE+$V_{E,\tau}$ interaction the overlap is negligible. It is particularly 
troubling that the TPE+$V_{E,\tau}$ interaction predicts negative pressures at 
$2n_0$. 

As argued earlier, these shorter-range contributions are an artifact of using local 
regulators. The large attractive contribution of the TPE+$V_{E,\tau}$ interaction 
is unphysical and should be discarded for the following reasons: (i) the full 
N$^2$LO 3N contributions in momentum-space calculations by \cite{Hebeler:2009iv, 
Tews:2012fj, Hagen:2013yba, Sammarruca:2014zia, Wlazlowski:2014jna}, and \cite{Drischler:2016djf} are all found to be repulsive in PNM; 
(ii)  the inclusion of N$^3$LO $3N$ contributions does not change the repulsive 
nature of $3N$ forces in PNM, see~\cite{Kruger:2013kua} and \cite{Drischler:2016djf}; 
and (iii) local regulators lead to less repulsion from the $3N$ TPE compared to 
typically used nonlocal regulators~\citep{Tews:2015ufa,Dyhdalo:2016ygz}. For 
these reasons, we will ignore the N$^2$LO TPE+$V_{E,\tau}$ interaction in the
following but include the TPE+$V_{E,\mathbbm{1}}$ interaction to investigate 
more repulsive short-range $3N$ forces.

It is worth noting that the pressure predicted by the phenomenological AV8'+UIX 
interaction at $2n_0$ is significantly larger than the  N$^2$LO (TPE-only) prediction. 
Even the inclusion of additional repulsion through the N$^2$LO 
TPE+$V_{E\mathbbm{1}}$ interaction does not alleviate this discrepancy. 
With some reserve, we propose that our calculations with the N$^2$LO TPE+$V_{E,\mathbbm{1}}$ interaction provide an upper limit to the pressure.
Calculations with local chiral interactions suggest that $P (2n_0) <  20$ MeV fm$^{-3}$ 
in PNM, see Table~\ref{tab:convergence} . 

In the following, we use the PNM results from the chiral N$^2$LO TPE-only and TPE+ $V_{E,\mathbbm{1}}$ interactions up to a transition density $n_{\text{tr}}$, which we vary in the range of $1-2 n_0$. Although the uncertainties are sizable at $2n_0$, we shall find that PNM calculations still provide useful constraints. The phenomenological AV8'+UIX interaction will also be analyzed in the same density interval for comparison.  Even though the expansion parameter $\simeq k_F/\Lambda_B$ only increases by about $25\%$ when the density is increased from $n_0$ to $2n_0$, we have chosen $2n_0$ as an upper limit to the validity of nuclear Hamiltonians for the following reasons. 
First, the previous discussion of uncertainties and the order-by-order convergence of the energy per particle and pressure in neutron matter has shown that while the convergence for the energies is consistent with EFT expectations, the situation is less satisfactory for the pressure at $2n_0$. 
Second, the accuracy of chiral nuclear interactions in describing typical momenta in nuclei and nucleon-nucleon scattering data decreases with increasing density. 
Third, at higher densities, additional degrees of freedom, e.g., hyperonic dof, might appear~(e.g.,~\cite{Ambartsumyan:1960, Glendenning:1982nc, Lonardoni:2014bwa, Gal:2016boi}).
Fourth, at densities above $2 n_0$, typical momenta in neutron matter are comparable to the cutoff scales employed in the calculation, which further increases the size of regulator artifacts. 
Based on these reasons, we believe that $2n_0$ is a reasonable upper bound for calculations with the local chiral Hamiltonians that we employ here.

\subsection{The EOS of neutron-star matter}

\begin{figure}[t]
\centering
\includegraphics[trim= 0 0 0 0, clip=,width=0.6\textwidth]{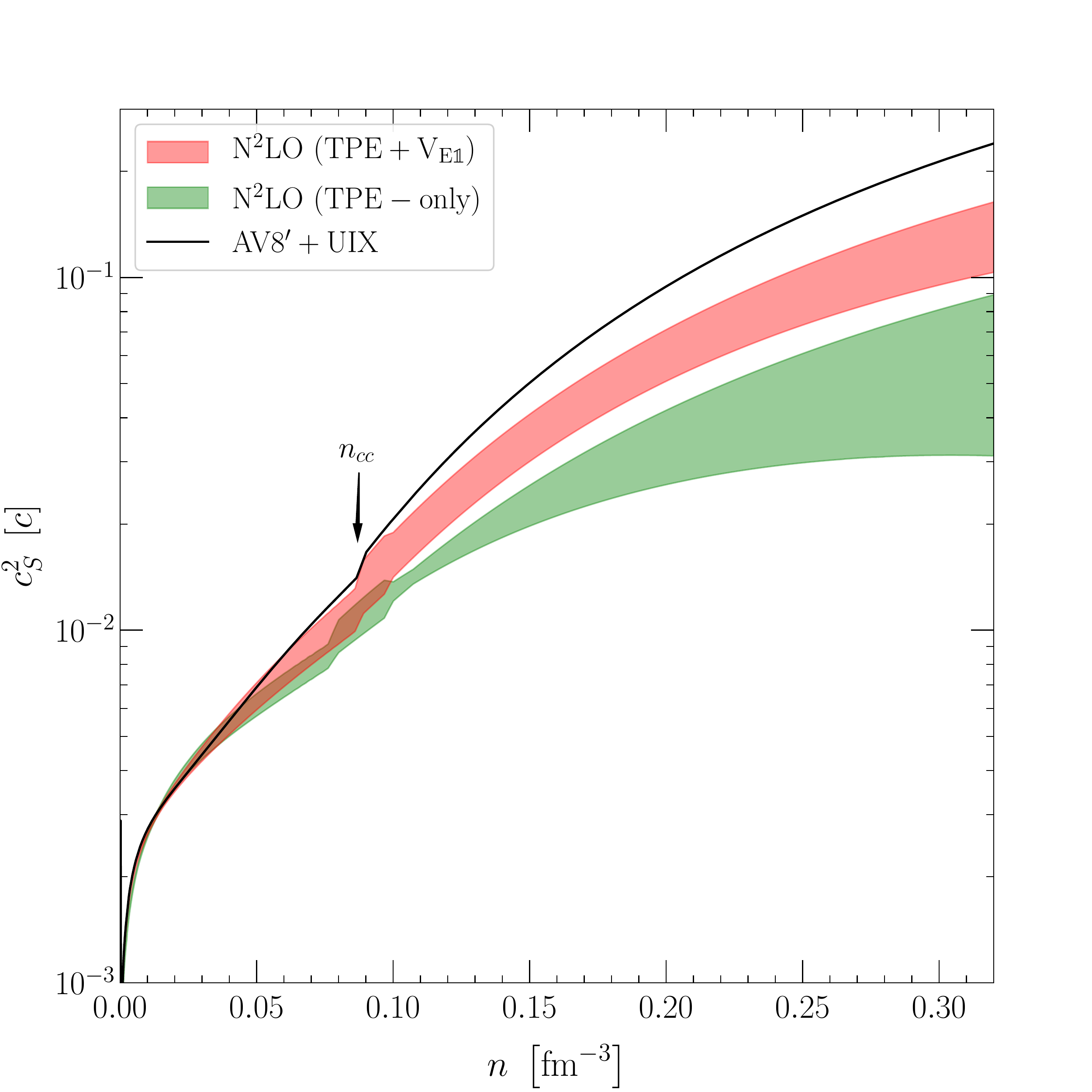} 
\caption{Speed of sound as a function of density for NS matter based 
on the local chiral N$^2$LO TPE-only (green lower band) and TPE+$V_{E,\mathbbm{1}}$
(red higher band) interactions of~\cite{Lynn:2015jua} and the AV8'+UIX interaction (black line) for
comparison. The arrow indicates the region of the crust-core transition.
\label{fig:cS_nuc}}
\end{figure}

Matter in NSs is in $\beta-$equilibrium, and at the relevant densities a small fraction of protons will be present. The proton fraction, denoted by $x$, increases with density but remains small and $x \lesssim 10\%$ even at $2n_0$. Although the proton contribution to the EOS can be expected to be small compared to the intrinsic uncertainty associated with the nuclear Hamiltonian discussed earlier, we shall extend the PNM results to finite proton fraction. To achieve this, we use the parameterization introduced by~\cite{Hebeler:2013}, given by
\begin{align} \label{eq:parametrization}
\frac{E^{\text{nuc}}}{A}(n,x)&= T_0  \left[\frac35\left(x^{\frac53}
+(1-x)^{\frac53} \right)\left(\frac{2n}{n_0}\right)^{\frac23}\right.- \left[(2\alpha - 4\alpha_L) x (1 - x) +  \alpha_L \right]\cdot \frac{n}{n_0} \\
&\quad \left. + \left[(2\eta - 4\eta_L) x (1 - x) +  \eta_L \right]\cdot \left(\frac{n}{n_0}\right)^{\gamma} \right] \nonumber\,,
\end{align}
where $T_0=(3 \pi^2 n_0/2)^{2/3}\hbar^2/(2 m_N)$ is the Fermi energy 
of noninteracting symmetric nuclear matter at saturation density, $x=n_p/n$ is the 
proton fraction ($n_p$ is the proton density), and $\alpha, \alpha_L, \eta,
\eta_L$, and $\gamma$ are parameters that are fit to the neutron-matter results 
and the saturation point of symmetric nuclear matter, 
$\frac{E^{\text{nuc}}}{A}(n_0,0.5)=-16 \mev$, and $P^{\text{nuc}}(n_0,0.5)=0$. 
The saturation point determines $\alpha$ and $\eta$, while the parameters 
$\alpha_L$, $\eta_L$, and $\gamma$ are determined by the PNM results. 
This parameterization provides a faithful reproduction of the PNM results 
obtained using AFDMC for densities up to $2n_0$, and has also been shown 
to provide a a good representation of results for asymmetric nuclear matter 
obtained in many-body perturbation theory~\citep{Drischler:2013iza}. 

Using the parameterization in Equation~\eqref{eq:parametrization}, we follow 
\cite{Tews:2016ofv} to construct a consistent crust model up to the
crust-core transition density, $n_{cc} \approx n_0/2$. For densities between 
$n_{cc}$ and the chosen transition density $n_{\text{tr}}$, we extend the 
PNM results to $\beta$ equilibrium. From this procedure, the NS equation 
of state 
$P(\epsilon)$ and the speed of sound $c_S^2(n)$ are determined. 
In Figure~\ref{fig:cS_nuc}, we show the speed of sound in NS matter 
up to two times nuclear 
saturation density for the chiral N$^2$LO TPE-only (green band), N$^2$LO 
TPE+$V_{E,\mathbbm{1}}$ (red band), and AV8'+UIX (black line) interactions. 

\section{Speed-of-sound extension to higher densities}\label{sec:EOSextension}

To obtain the mass-radius relation of NSs, we need to extend 
our EOS of NS matter to higher densities. A common approach is
to use a polytropic extension (see, e.g., \cite{Hebeler:2013}, 
\cite{Kurkela:2014vha}, and \cite{Raithel:2016bux} for more details). In such an 
approach, the higher-density EOS is parameterized by a set of piecewise 
polytropes, that are matched to the microscopic calculations. The polytropic 
indices and the transition densities between the individual segments are then
varied to sample many possible EOS curves. This approach is rather general 
but it leads to discontinuities in the speed of sound. 

\begin{figure}[t]
\centering
\includegraphics[trim= 0 0 0 0cm, clip=,width=0.6\textwidth]{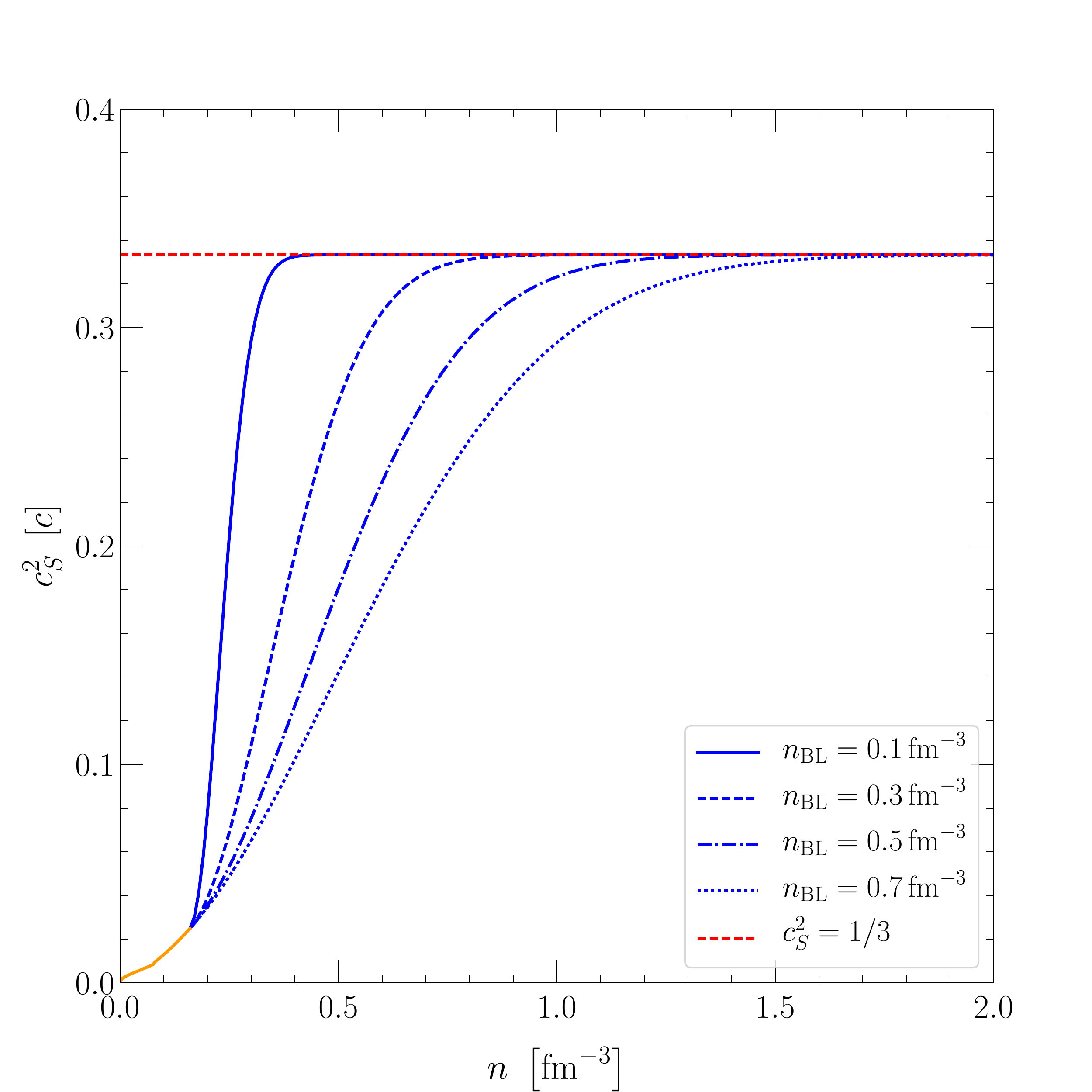} 
\caption{Extension for $c_S^2$ as a function of density $n$ as defined in 
Equation~\eqref{eq:baseline}. The orange line represents the constraint from the 
TPE-only EOS (up to $n_{\text{tr}}=0.16 \fmiq$), while the blue lines 
are extensions for several widths ($n_{\rm BL}=0.1, 0.3, 0.5, 0.7 \fmiq$ from 
left to right). The red dashed line indicates the conformal limit.
\label{fig:cS2_baseline}}
\end{figure}

In this work, we shall restrict our analysis to scenarios for which the speed of 
sound is continuous for densities encountered inside NSs; allowing us 
to directly parameterize the speed of sound and use it to construct the EOS. 
Although this may be less general than EOSs constructed from piecewise 
polytropes, our choice is motivated by the following observation. Our 
calculations of the nuclear EOS up to $2n_0$ show that it is relatively soft with 
a rather small speed of sound. To obtain a maximum NS mass $M_{\rm max} > 2$ 
$M_\odot$, the EOS at higher density needs to stiffen significantly. This 
disfavors strong first-order phase transitions inside NS above $2n_0$, and models of high-density matter where new Fermionic or bosonic 
degrees of freedom appear suddenly to produce discontinuities in the energy 
density (note that the pressure is continuous and monotonically increasing 
towards the center of NS due to hydrostatic equilibrium; \cite{Alford:2013aca}). 
Without such phase transitions, it is more natural to expect that the evolution 
of the speed of sound in the NS core will be continuous. In the following, we 
shall connect the speed of sound in neutron matter at densities up to $1-2 n_0$ 
shown in Figure~\ref{fig:cS_nuc}, to the speed of sound expected in deconfined  
quark matter at very high density~\citep{Kurkela:2009gj} using two 
parameterizations, that we will discuss in Sections~\ref{sec:csconformal} and \ref{sec:csnonconf}. 

We stress that sampling the speed of sound is a complimentary approach to using a set of piecewise polytropes. In fact, choosing a set of piecewise polytropes is completely equivalent to choosing piecewise segments for the speed of sound that have a specific functional form and are connected by phase transitions, e.g., three segments with three such phase transitions in the case of~\cite{Hebeler:2013}. Sampling the speed of sound allows us to easily control the number of phase transitions as well as their characteristics and to extract meaningful information on its density behavior, e.g., the maximum speed of sound inside an NS and its peak position. It is also straightforward to include additional information on the speed of sound, e.g., bounds on $c_S$. In Section~\ref{sec:csconformal}, we will analyze the impact of such a bound.

\begin{figure*}[t]
\centering
\includegraphics[trim= 0 0 0 0, clip=,width=0.49\textwidth]{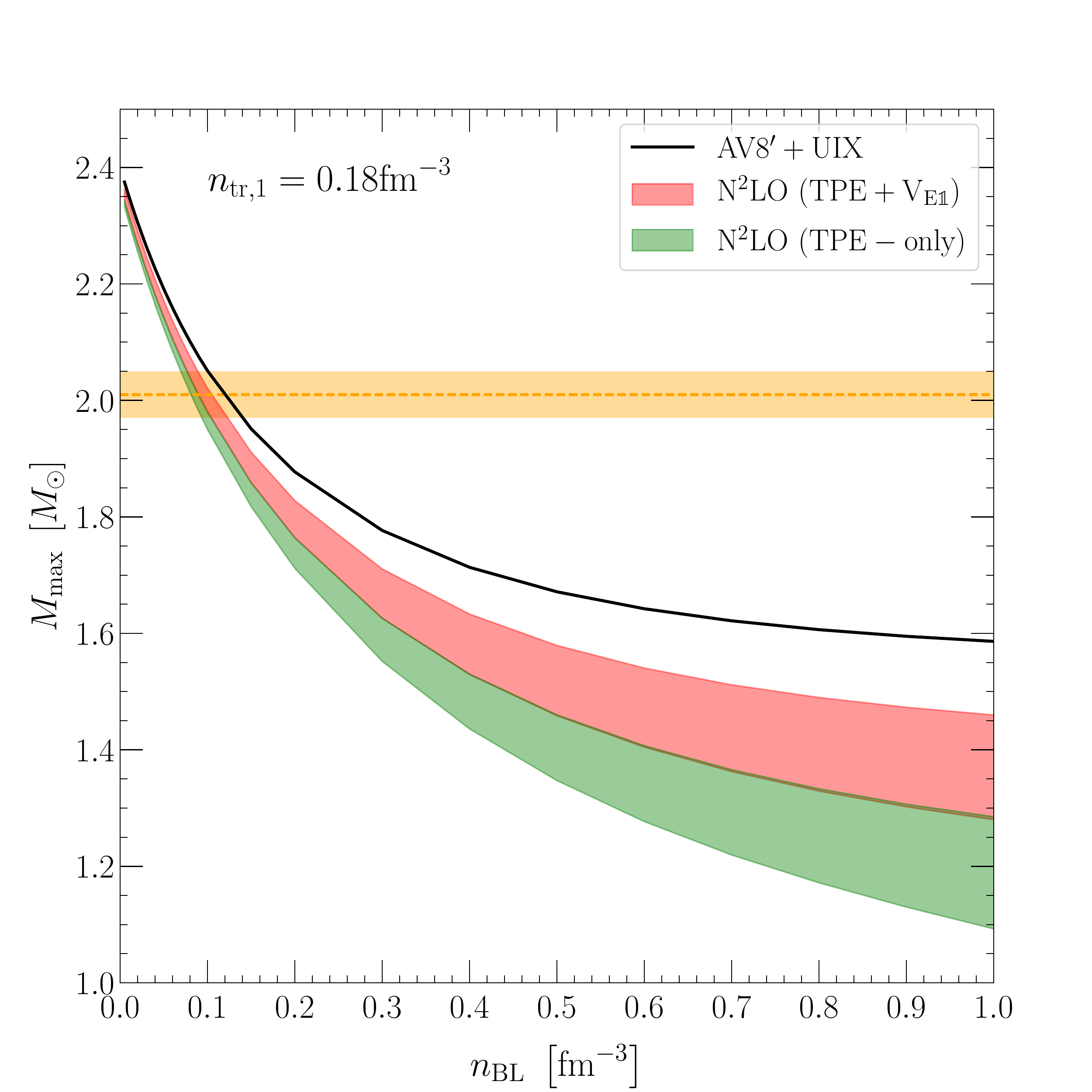} 
\includegraphics[trim= 0 0 0 0, clip=,width=0.49\textwidth]{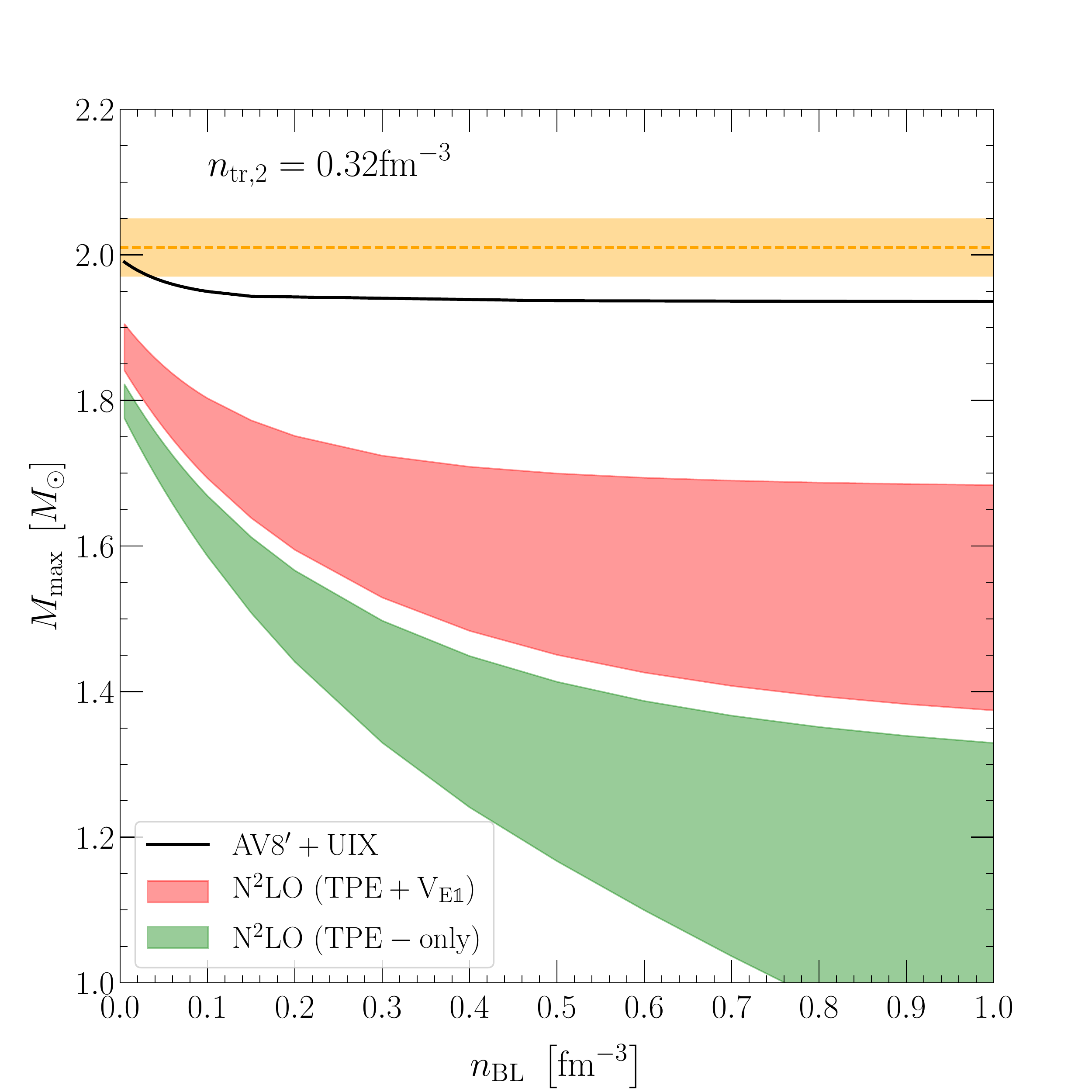} 
\caption{Maximum mass $M_{\rm max}$ as a function of baseline 
width $n_{\rm BL}$ of Equation~\eqref{eq:baseline} for the chiral N$^2$LO 
interactions (red and green bands), and the AV8'+UIX interaction (black line) for 
two transition densities, $n_{\text{tr,1}}=0.18 \fmiq$ and 
$n_{\text{tr,2}}=0.32 \fmiq$. We also indicate the highest observed 
NS mass with its uncertainty (orange band).
\label{fig:Mvsw}}
\end{figure*}

\subsection{Extension with the Conformal Limit}\label{sec:csconformal}

In this scenario $c_S^2<1/3$ at all densities, and we extend the speed of sound
to densities above those described by the nuclear EOS using a simple 
three-parameter curve, 
\begin{align}
c_S^2=\frac13-c_1\exp \left[-\frac{(n-c_2)^2}{n_{\rm BL}^2} \right]\,,\label{eq:baseline}
\end{align}
where two of the parameters, $c_1$ and $c_2$, are fit to the speed of sound 
and its derivative at $n_{\text{tr}}$. The remaining parameter, $n_{\rm BL}$, 
controls the width of the curve and presents a density interval in which the speed
of sound changes considerably. The smaller this parameter, the stronger the 
change of the speed of sound with density. By varying this remaining parameter, 
we can easily control the slope of the speed of sound after the transition density, 
see Figure~\ref{fig:cS2_baseline}. The choice of an exponential function may 
seem ad hoc at this stage, but will be well motivated shortly when we find that to 
obtain $2$ M$_\odot$ NSs,  $c_S$ must increase rapidly over a narrow interval $n_{\rm BL} \lesssim n_0$.  

From Eq.~\ref{eq:baseline}, we can construct the EOS using the following procedure.
Starting at $n_{\text{tr}}$, where $\epsilon(n_{\text{tr}})$,  
$p(n_{\text{tr}})$, and $\epsilon'(n_{\text{tr}})$ are known, we take successive small 
steps $\Delta n$ in density, 
\begin{align}
n_{i+1}&= n_i + \Delta n \\
\epsilon_{i+1} &= \epsilon_i +\Delta\epsilon= \epsilon_i + \Delta n \cdot \left(\frac{\epsilon_i+p_i}{n_i}\right) \\
p_{i+1} &= p_i + c_S^2 (n_i) \cdot \Delta \epsilon\,,
\end{align}
to iteratively obtain the high-density EOS where the index $i=0$ defines the transition density $n_{\text{tr}}$. Note, in the second line 
we have used the thermodynamic relation $p=n \partial \epsilon/\partial n -\epsilon$ valid at zero temperature. We then use the 
resulting equation of state to solve the Tolman-Oppenheimer-Volkoff  (TOV) 
equations, and to obtain the mass-radius relation for NSs.

\begin{figure*}[t]
\centering
\includegraphics[trim= 0 0 0 0, clip=,width=0.49\textwidth]{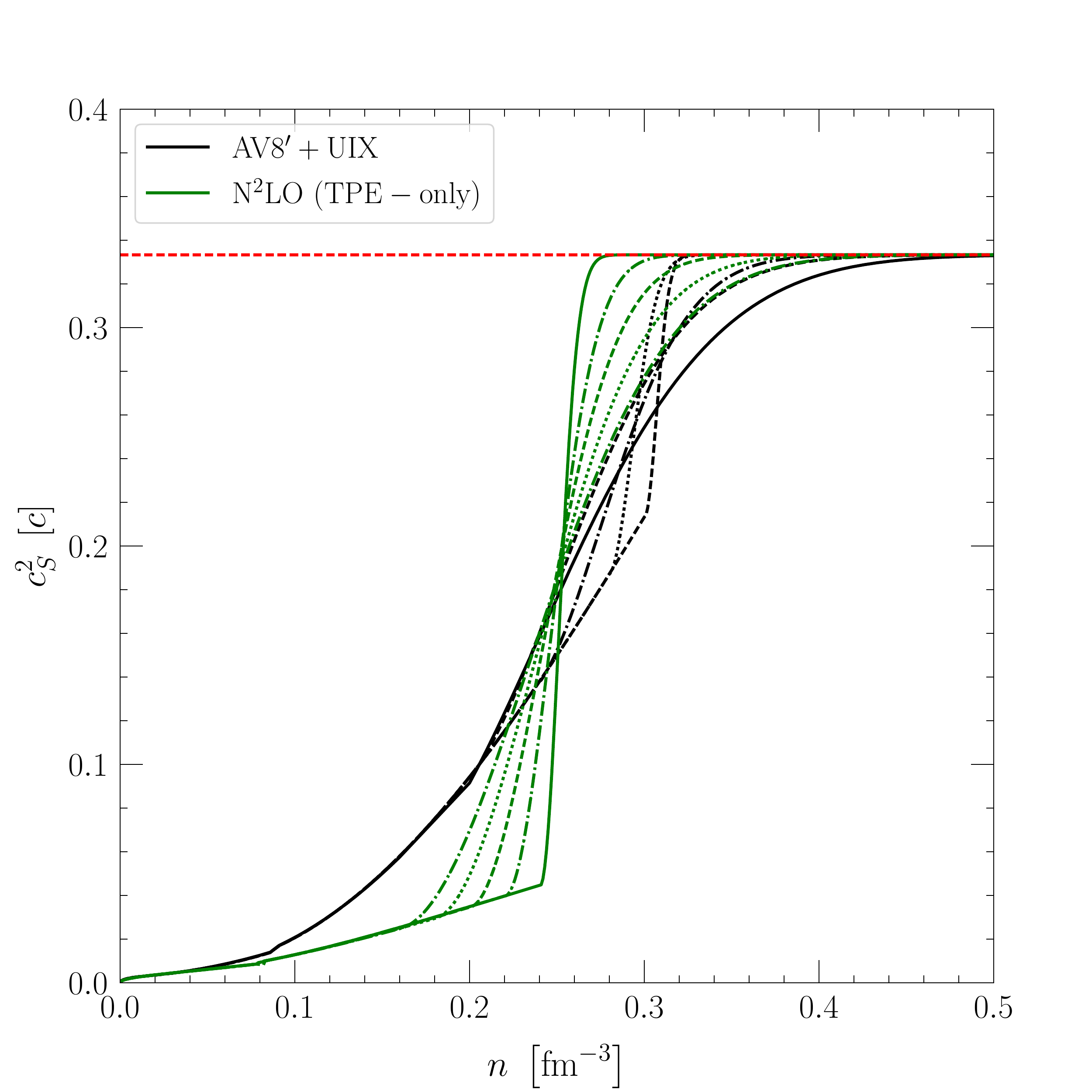} 
\includegraphics[trim= 0 0 0 0, clip=,width=0.49\textwidth]{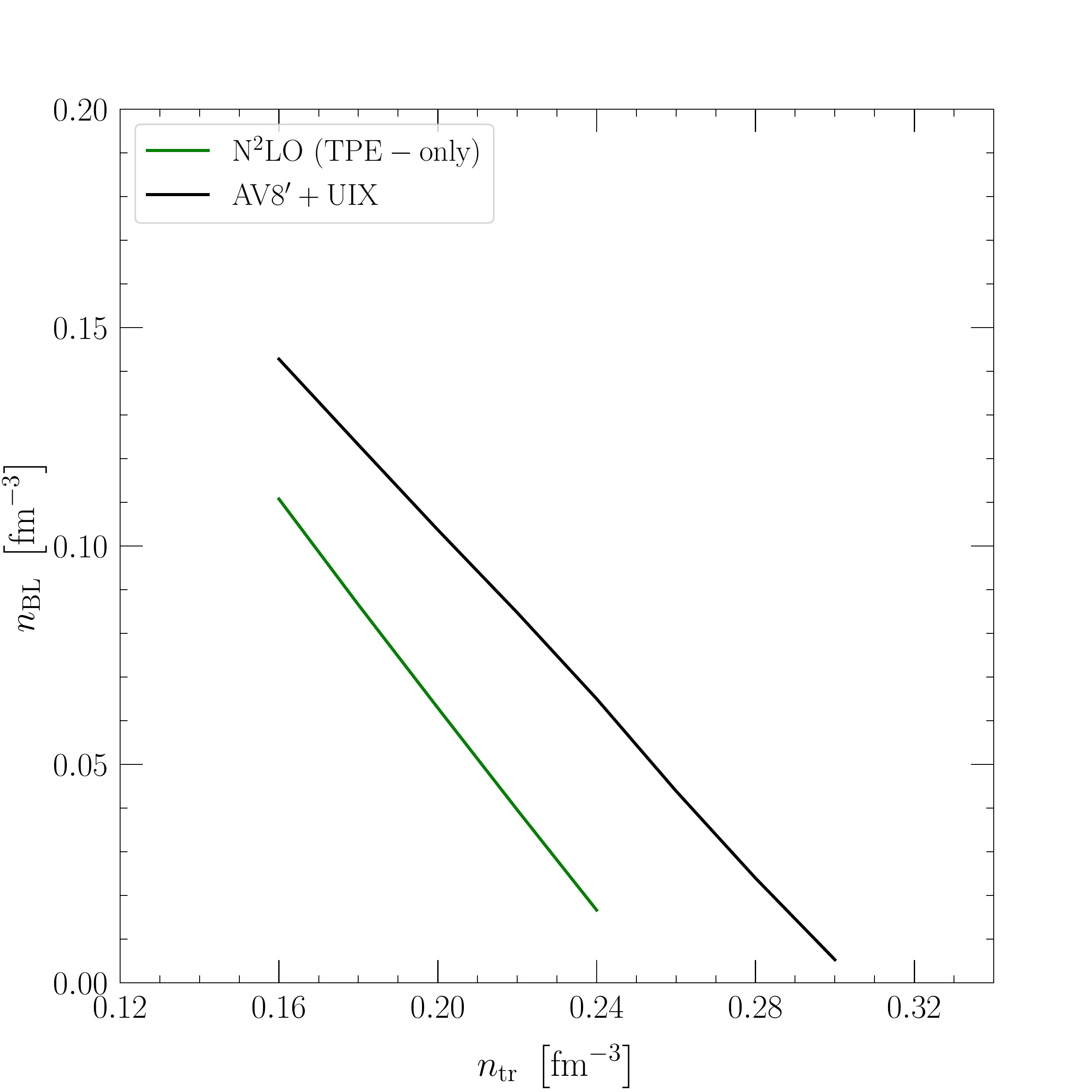} 
\caption{Left panel: $c_S^2$ as a function of density for several transition 
densities and two nuclear Hamiltonians. For each transition density and 
Hamiltonian, we show the curve for the maximal $n_{\rm BL}$ that still 
supports a two-solar-mass NS. Right panel: the maximal $n_{\rm BL}$ that is sufficient to support a two-solar-mass 
NS as a function of transition density for the same lines.
\label{fig:Cs_crit}}
\end{figure*}

In Figure~\ref{fig:Mvsw}, we show the maximum NS mass as a 
function of the baseline width $n_{\rm BL}$ of Equation~\eqref{eq:baseline} 
(describing the slope) for two transition densities, $n_{\text{tr,1}}=1.1 n_0
=0.18 \fmiq$ and $n_{\text{tr,2}}=2 n_0=0.32 \fmiq$, and for the 
nuclear interactions of Figure~\ref{fig:cS_nuc}. We compare these to the current 
NS maximum-mass constraint of $2.01\pm 0.04 $ M$_\odot$ 
\citep{Antoniadis2013}. Depending on the transition density and the nuclear
Hamiltonian, we observe the tension between current nuclear and astrophysical 
constraints and the $c_S^2 < 1/3$ bound set by the conformal limit, first noted in  
\cite{Bedaque:2014sqa} (see also \cite{Moustakidis:2016sab}). At the lower 
transition density, there is some freedom
to find an EOS that satisfies both the conformal bound and the maximum-mass 
constraint. However, the inferred value of $n_{\rm BL}$ is small, indicating that 
$c_S$ must increase much more rapidly in the density interval $n_0-2n_0$ than 
would be compatible with EFT predictions. For $n_{\text{tr}}=2n_0$, chiral EFT, 
even with the repulsion beyond the TPE-only interaction, is unable to support the 
$2.01\pm 0.04 $ M$_\odot$ NS mass if the conformal bound is observed. 
The stiffest EOS predicted by the phenomenological Hamiltonian AV8'+UIX 
(black line) barely satisfies both bounds. 

In the left panel of Figure~\ref{fig:Cs_crit}, we show the evolution of $c^2_S$ for 
the largest values of $n_{\rm BL}$ compatible with a two-solar-mass NS
for several transition densities and for the softest and stiffest nuclear interaction 
under investigation (without uncertainty bands for clarity). In the right panel, we 
show this maximal width as a function of $n_{\rm tr}$. We find that for the chiral 
N$^2$LO TPE-only interaction there is no curve that satisfies both the 
speed-of-sound and the NS mass bounds for transition densities larger 
than $\approx 0.25 \fmiq$.  For the phenomenological nuclear Hamiltonian, the 
situation is only slightly different. We find that for low transition densities a moderate 
slope of the speed of sound is sufficient to reproduce a two-solar-mass NS. With increasing transition density, again, the speed of sound has to increase 
on a smaller density interval to fulfill the mass constraint. Above transition densities 
of $\approx 0.3 \fmiq$, we find strong tension between both bounds.

To summarize, since AFDMC predictions for the EOS using EFT interactions are 
very reliable, and order-by-order convergence at these densities was observed, we
argue that our results suggest that either QCD violates the conformal bound, 
or a chiral EFT based description breaks down at densities below $2 n_0$ 
due to new physical effects.

\subsection{Extension without the Conformal Limit}\label{sec:csnonconf}

\begin{figure}[t]
\centering
\includegraphics[trim= 0 0 0 0cm, clip=,width=0.6\textwidth]{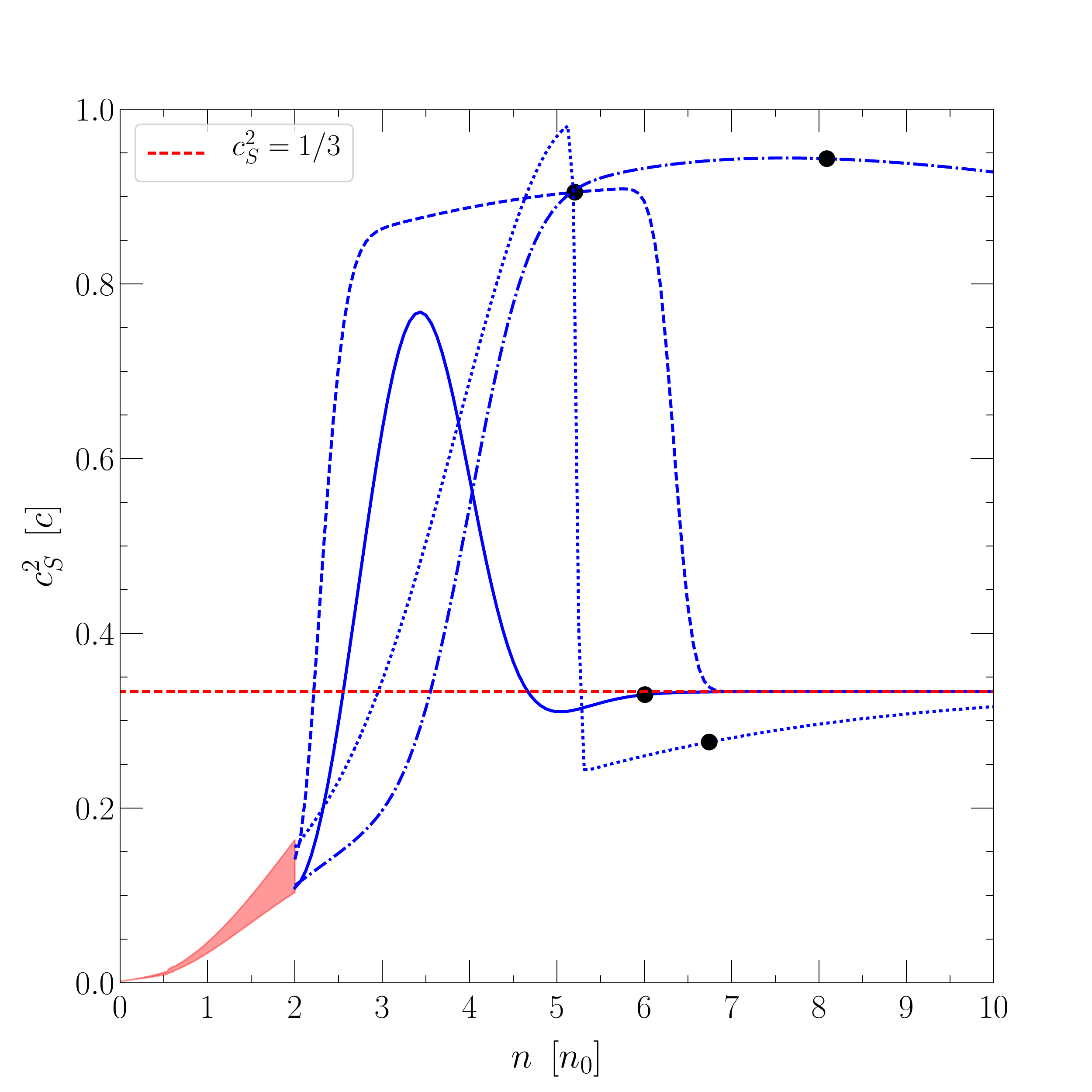} 
\caption{Four examples for the extension of $c_S^2(n)$ as defined in 
Equation~\eqref{eq:Csmodel} (without conformal limit) for $n_{\rm tr,2}$ and the 
chiral TPE+$V_{E,1}$ interaction. Black dots indicate the maximal central 
densities reached inside NSs for the corresponding EOSs and the 
red dashed line indicates the conformal limit.
\label{fig:cS2_curves}}
\end{figure}

\begin{figure*}[t]
\begin{center}
\includegraphics[trim= 0 0 0 0cm, clip=,width=0.285\textwidth]{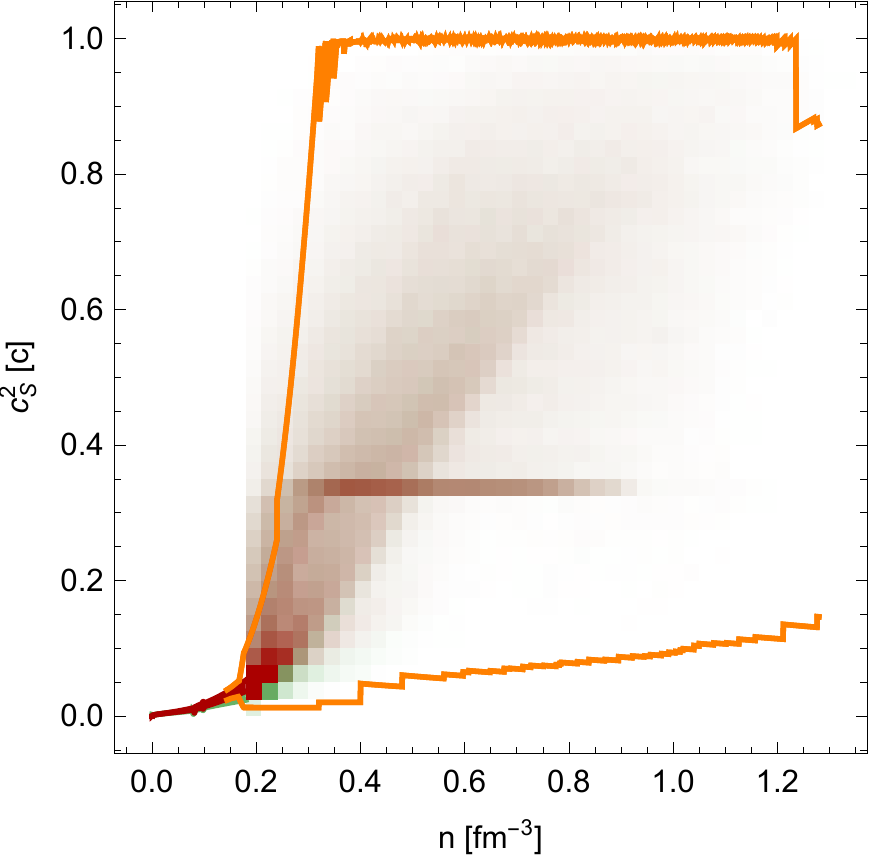} \hspace*{0.2cm}
\includegraphics[trim= 0 0 0 0cm, clip=,width=0.285\textwidth]{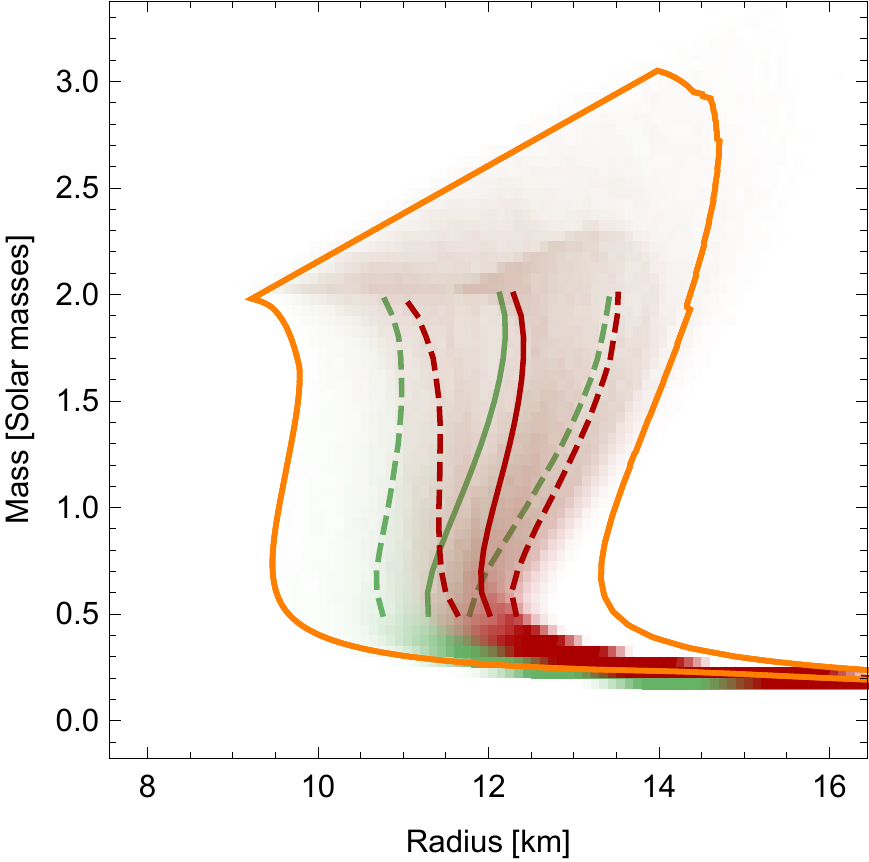} \hspace*{0.2cm}
\includegraphics[trim= 0 0 0 0cm, clip=,width=0.285\textwidth]{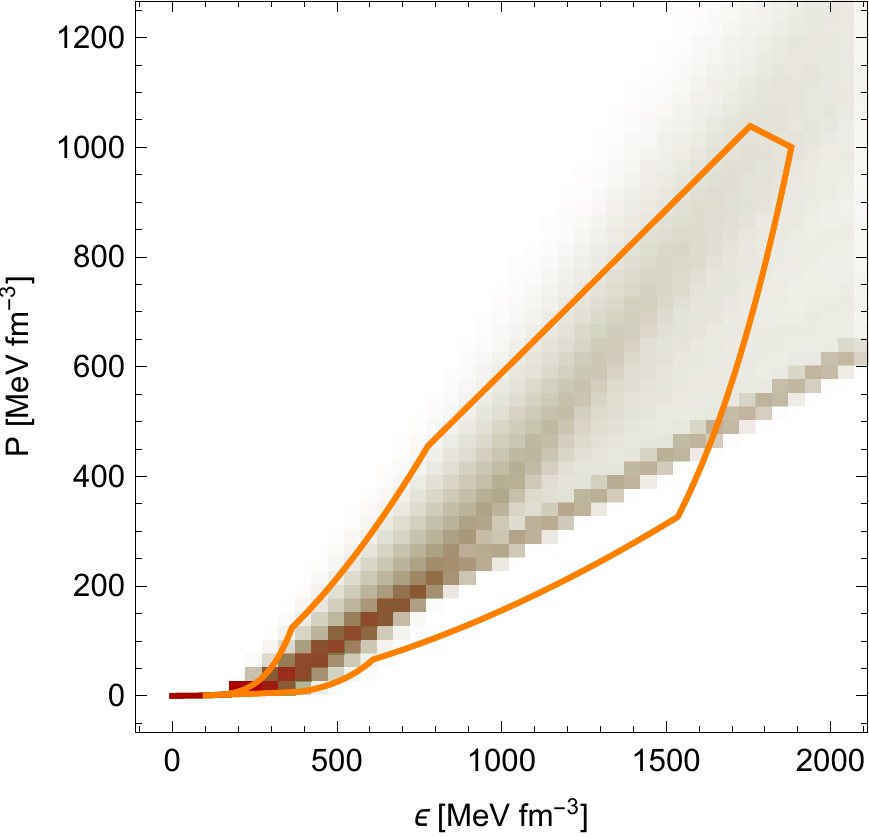} \\
\vspace*{0.2cm}
\includegraphics[trim= 0 0 0 0cm, clip=,width=0.285\textwidth]{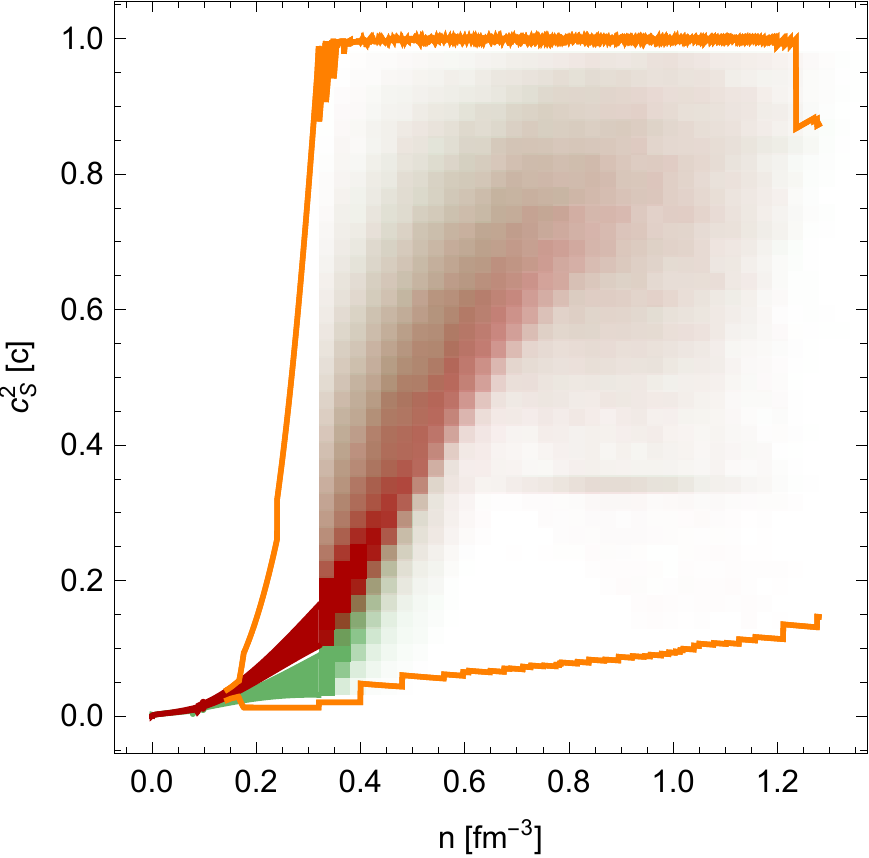} \hspace*{0.2cm}
\includegraphics[trim= 0 0 0 0cm, clip=,width=0.285\textwidth]{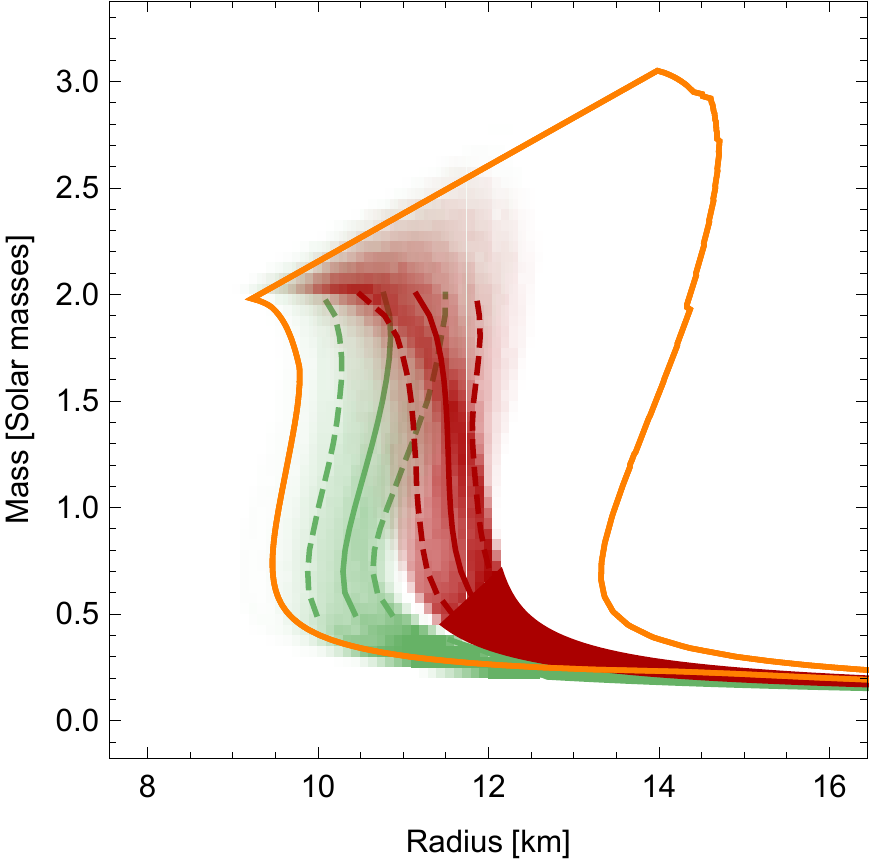} \hspace*{0.2cm}
\includegraphics[trim= 0 0 0 0cm, clip=,width=0.285\textwidth]{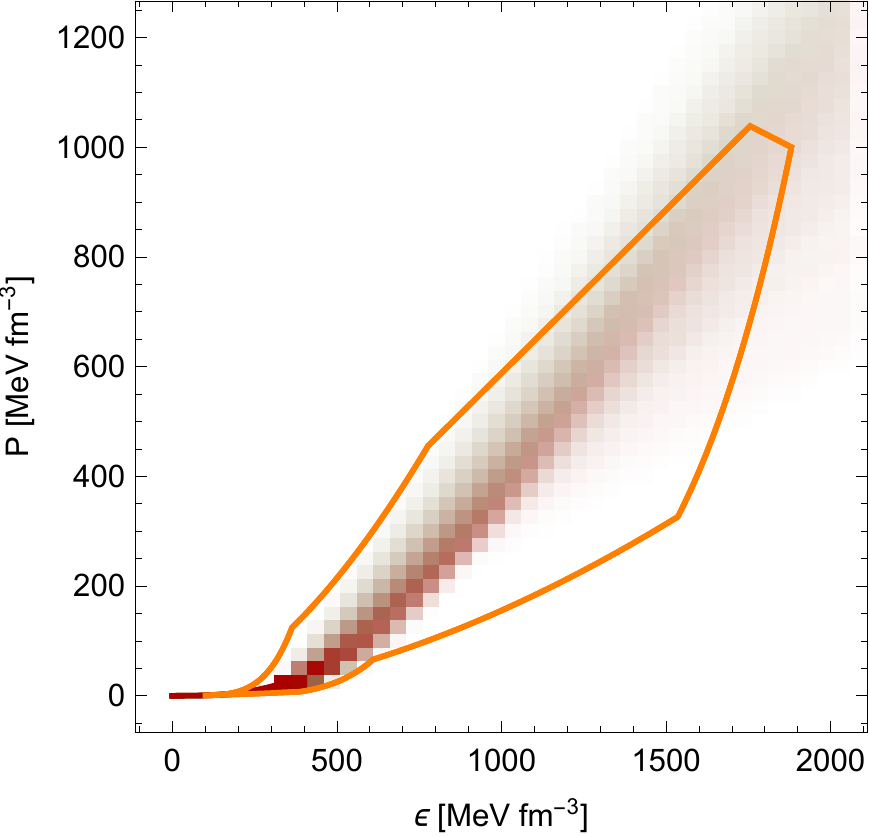} 
\end{center}
\caption{\label{fig:2DHistogramTPE}
Histograms for $c_S^2(n)$, the mass-radius relation, and the EOS for all the 
accepted parameter sets for the local chiral N$^2$LO interactions of 
Figure~\ref{fig:cS_nuc} and $n_{tr,1}$ (upper panels) and $n_{tr,2}$ (lower panels). 
For the $c_S^2(n)$ histogram, we terminate each parameterization at its maximal 
central density. The orange lines are the corresponding contours for the polytropic 
expansion of~\cite{Hebeler:2013}. For the mass-radius curve, we also show the
average radius for each mass (solid line) as well as 68\% confidence intervals 
(dashed lines).} 
\end{figure*}

\begin{figure*}[th]
\begin{center}
\includegraphics[trim= 0 0 0 0cm, clip=,width=0.285\textwidth]{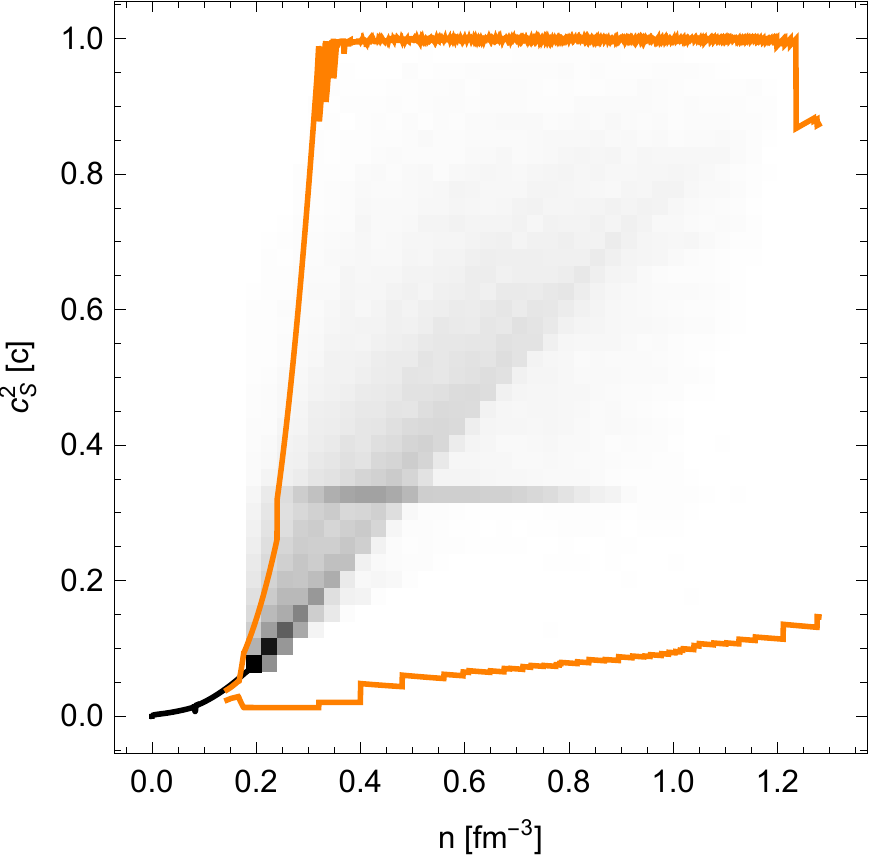} \hspace*{0.2cm}
\includegraphics[trim= 0 0 0 0cm, clip=,width=0.285\textwidth]{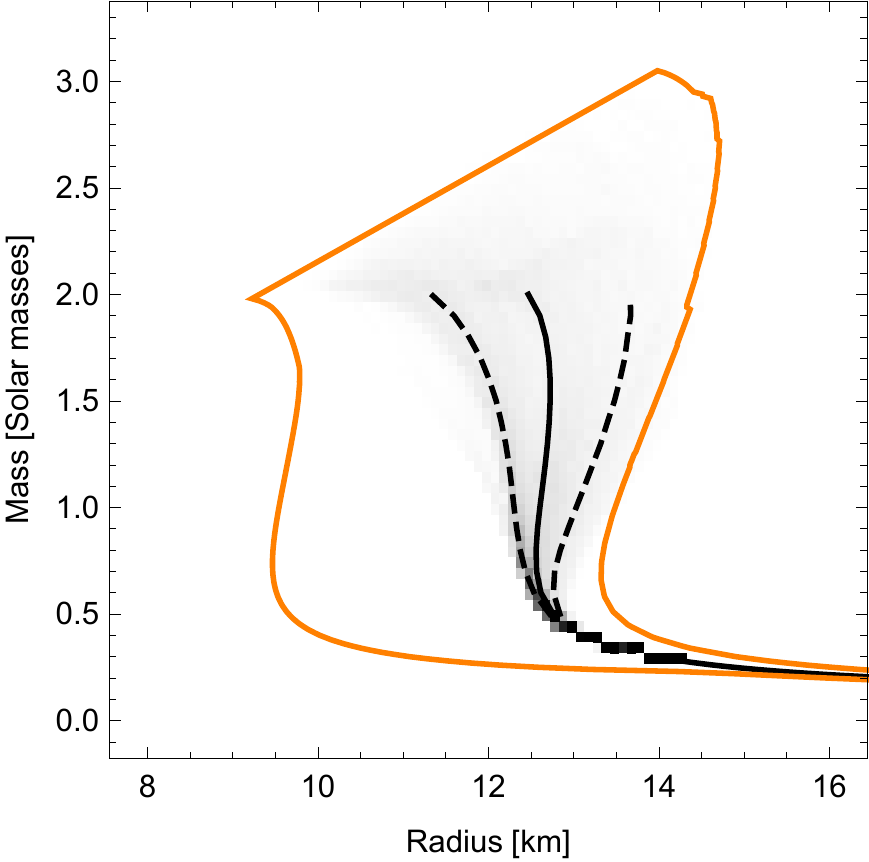} \hspace*{0.2cm}
\includegraphics[trim= 0 0 0 0cm, clip=,width=0.285\textwidth]{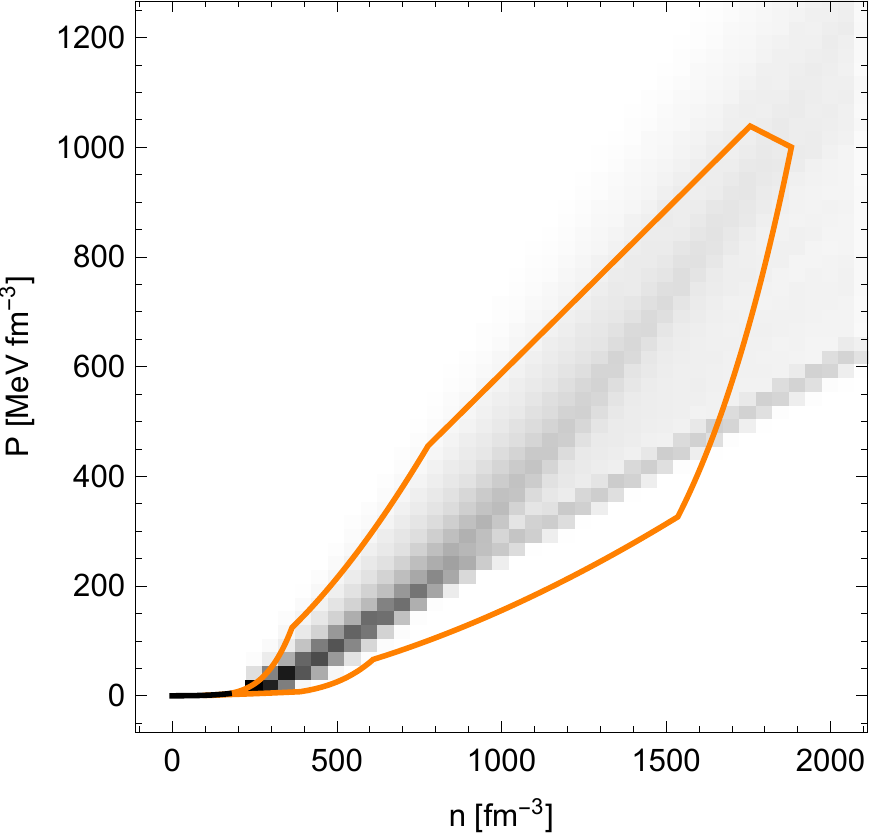} \\
\vspace*{0.2cm}
\includegraphics[trim= 0 0 0 0cm, clip=,width=0.285\textwidth]{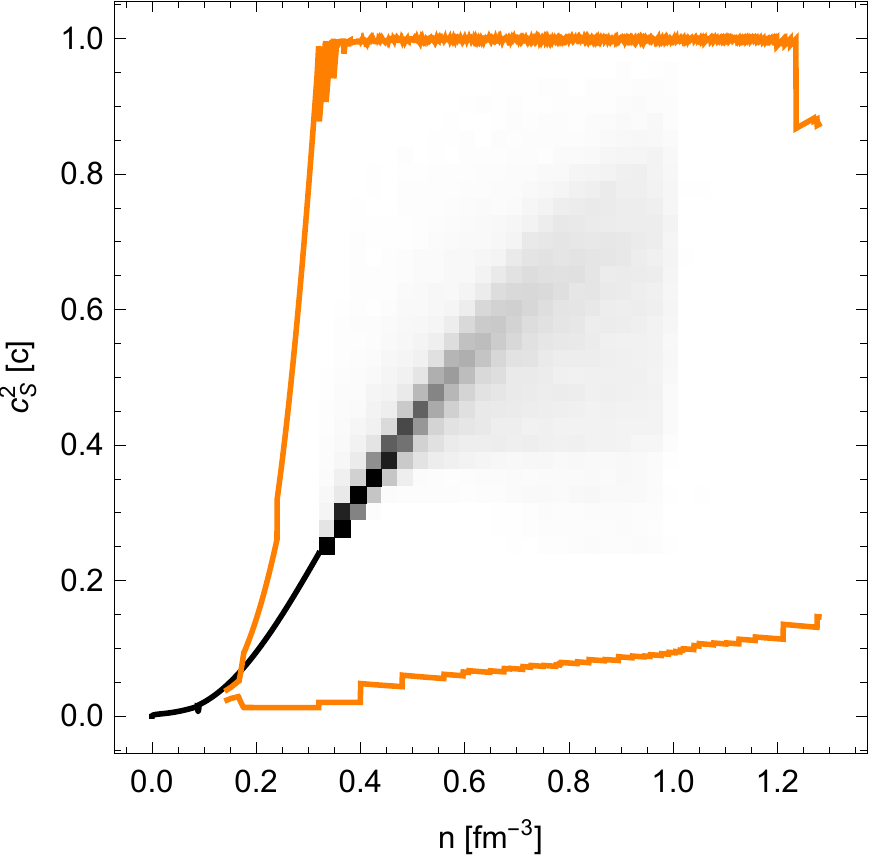} \hspace*{0.2cm}
\includegraphics[trim= 0 0 0 0cm, clip=,width=0.285\textwidth]{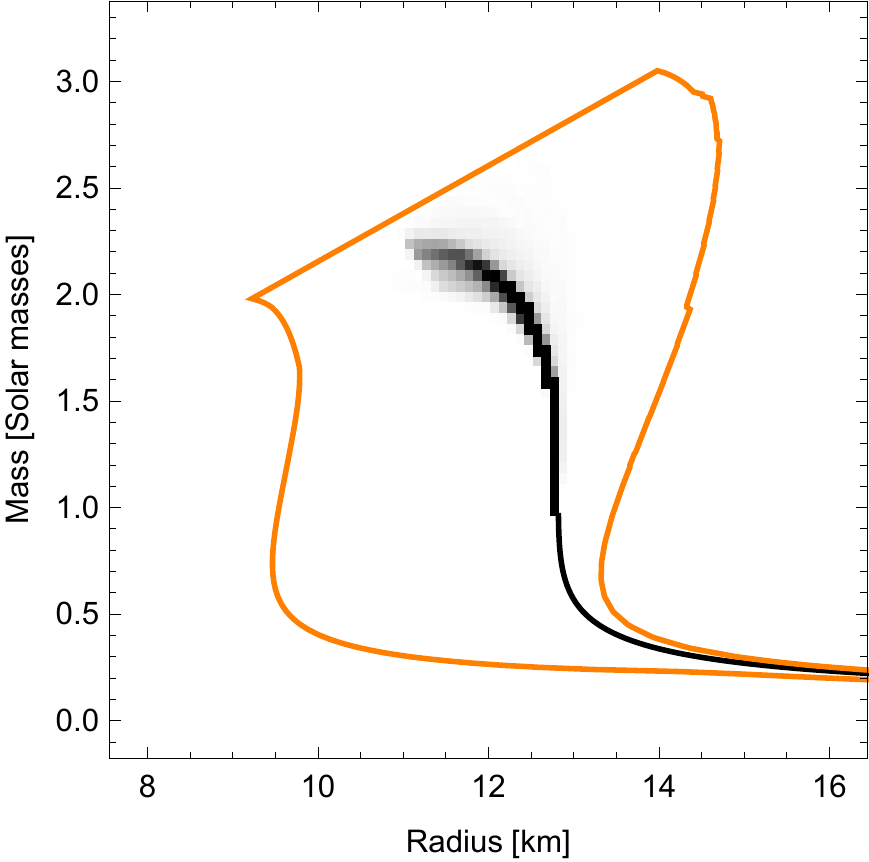} \hspace*{0.2cm}
\includegraphics[trim= 0 0 0 0cm, clip=,width=0.285\textwidth]{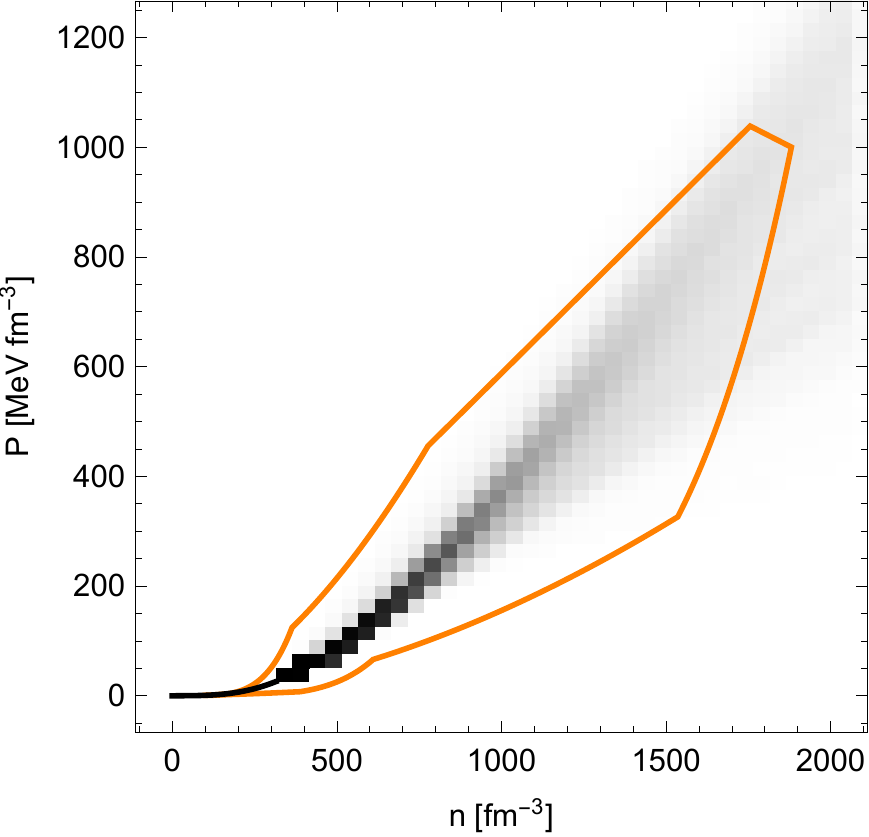} 
\end{center}
\caption{\label{fig:2DHistogramAV8UIX}
Same as Figure.~\ref{fig:2DHistogramTPE} but for the AV8'+UIX interaction 
(black line in Figure.~\ref{fig:cS_nuc}).} 
\end{figure*}

We shall now allow the speed of sound to exceed the conformal limit of 
$1/\sqrt{3}$ at the densities encountered in the NS core and only 
require that $c_S<1$. We model its evolution with density in a minimalist 
approach constrained by knowledge of its behavior at low and high density 
by adding a skewed Gaussian to the form defined in Equation~\eqref{eq:baseline}, 
\begin{align}
\label{eq:Csmodel}
c_S^2 &=\frac13-c_1\exp \left[-\frac{(n-c_2)^2}{n_{\text{BL}}^2} \right] +h_{\rm P} \exp{\left[-\frac{(n-n_{\rm P})^2}{w_{\text{P}}^2}\right]} 
\left(1+\rm{erf}\left[s_{\rm P}\frac{(n-n_{\rm P})}{w_{\text{P}}}  \right] \right)\,, 
\end{align}
where the peak is described by its height $h_{\rm P}$, position $n_{\rm P}$,
width $w_{\rm P}$, and the shape or skewness parameter $s_{\rm P}$. 
The coefficients $c_1$ and $c_2$ are again adjusted to the microscopic 
calculations up to the transition density $n_{\rm tr}$ to make sure that the 
speed of sound and its derivative are continuous. By varying the remaining 
parameters, we can model many possible curves for the speed of sound.  
In Figure~\ref{fig:cS2_curves}, we show four such parameterizations where we 
also indicate the maximal central density reached for each EOS. As one can see, 
this approach is very versatile and can produce very different shapes for the 
speed of sound, some of which resemble the result by \cite{Paeng:2016vpr}. 

The chosen form in Equation \ref{eq:Csmodel} allows $c_S$ to be a nonmonotonic 
function of density. The peak at intermediate density can be interpreted as 
describing a crossover transition that may be realized inside NSs. 
However, the high-density behavior, where $c_S^2\approx 1/3$, is realized 
at densities well beyond those accessed in NSs.  While this functional 
form is rather simple and has only five parameters, it is comparable in complexity 
to a polytropic extension with three polytropic segments, and can be easily 
extended. We have also tested different functional forms for the speed of sound 
and found our results to be robust: the average radius for a typical $1.4$ M$_\odot$ 
NS varied only on the percent level when different functional forms were chosen.

We sample values for the baseline width and the four peak parameters that 
determine $c_S$ to construct the high-density EOS. We use the full EOS and 
solve the TOV equations to obtain NS mass-radius curves and the NS maximum 
mass; when this is found to be greater than the $2.01 $ M$_\odot$ mass constraint, 
we accept the parameter set, and reject it otherwise. We sample the five 
parameters from uniform distributions within the following ranges: 
$n_{\rm BL}$ between $0.01-3.0 \fmiq$, 
$h_{\rm P}$ between $0.0-0.9 $, 
$w_{\rm P}$ between $0.1-5.0 \fmiq$, 
$n_{\rm P}$ between $(n_{\rm tr}+0.08)-5.0 \fmiq$,
and $s_{\rm P}$ between $(-50)-50$, and enforce that $0\le c_S^2 \le 1$. 

For the chiral interactions, we additionally sample from the uncertainty bands
for $c_S$ by randomly choosing a factor $f^{\rm err}$ between $-1$ and $1$, 
that linearly interpolates between the lower and upper bounds of the uncertainty 
band,
\begin{equation}
c_S(n)=c_S^{{\rm N}^2{\rm LO}}(n)+ f^{\rm err} \Delta c_S^{\rm EKM}(n)\,,
\end{equation}
where $c_S^{{\rm N}^2{\rm LO}}(n)$ is the chiral result at N$^2$LO and 
$\Delta c_S^{\rm EKM}(n)$ is its uncertainty. 

As stated before, we analyze the results for two different transition densities 
and generate a few thousand accepted parameter sets for each transition 
density. We show histograms 
for the resulting speed of sound, the mass-radius relation, and the EOS in
Figures~\ref{fig:2DHistogramTPE} and \ref{fig:2DHistogramAV8UIX} for both 
transition densities. For the mass-radius histograms, we also show the average
radius for each mass as well as 68\% confidence intervals.

We find that the speed of sound increases rapidly in a small density range 
above $n_{\rm tr}$. This increase is more drastic for softer nuclear interactions. 
For stiffer interactions, $c_S$ increases slowly and peaks at higher
densities. In all cases, for a large fraction of parameterizations, the speed of 
sound increases to values around 
$c_S\approx 0.9$. For the smaller transition density, there exist parameterizations 
that observe the conformal limit at all densities, while for the higher transition 
density all parameterizations violate this bound, consistent with our previous findings.

\begin{figure}[t]
\centering
\includegraphics[trim= 0 0 0 0cm, clip=,width=0.6\textwidth]{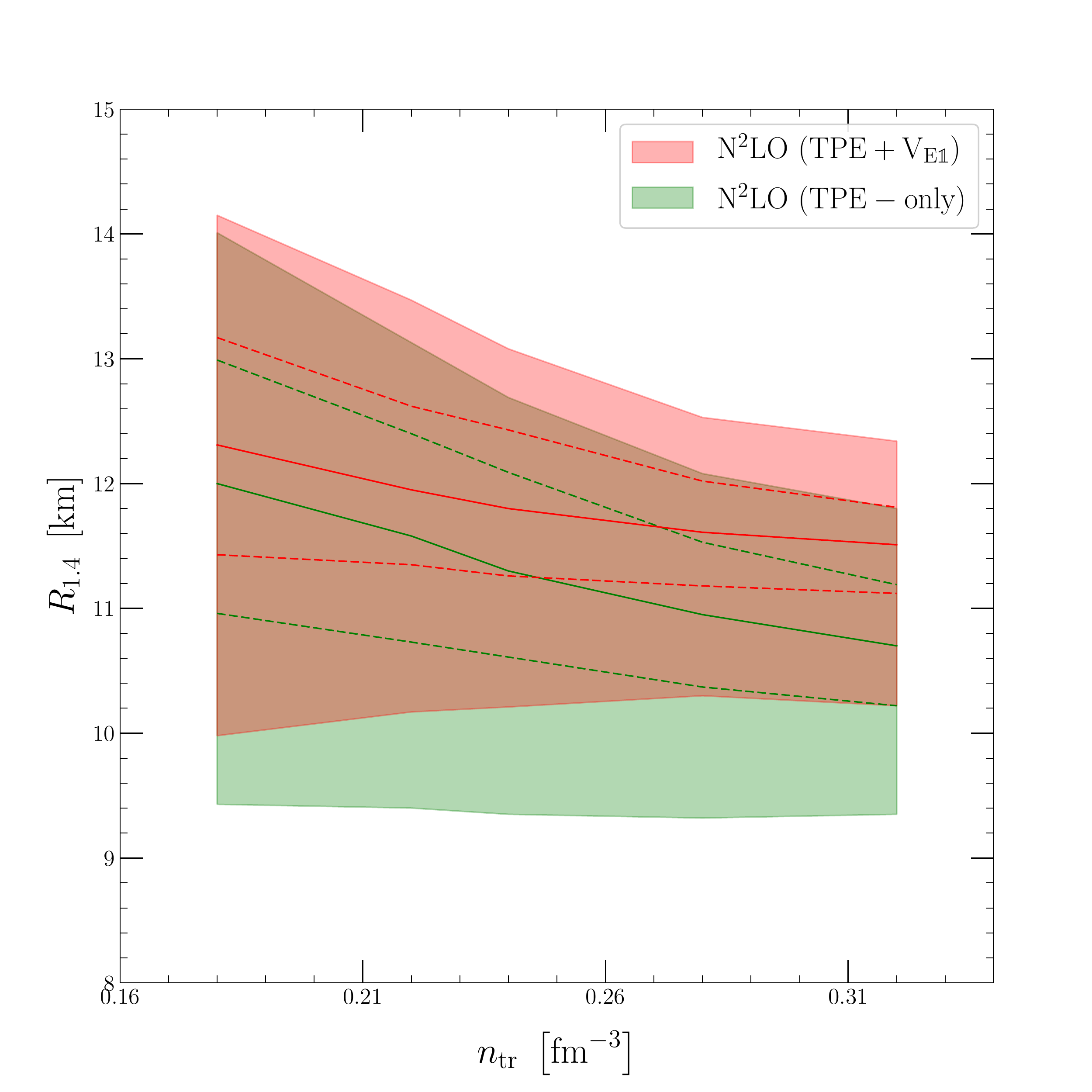} 
\caption{Radius of a $1.4$ M$_\odot$ NS as a function of transition 
density for the chiral models. We show the total range of models (colored bands)
as well as the mean (solid lines) with 68 \% intervals (dashed lines).
\label{fig:ntrRminRmax}}
\end{figure}

For the mass-radius relation, we find a rather broad radius distribution at 
lower transition densities that narrows with increasing transition density. This 
highlights the fact that PNM calculations at densities $\sim 2n_0$ provide 
valuable information despite sizable uncertainties. We highlight this fact in
Figure~\ref{fig:ntrRminRmax}, where we show the radius of a typical 
$1.4$ M$_{\odot}$ NS as a function of $n_{\rm tr}$ for the chiral EFT interactions.
At $n_{\rm tr,1}$, we find a radius range of $9.4-14.0$ km ($10.0-14.1$ km) with
a 68\% confidence interval of $12.0\pm1.0$ km ($12.3\pm0.9$ km) for the TPE-only
(TPE+$V_{E,\mathbbm{1}}$) interaction. This range reduces to $9.4-11.8$ km
($10.2-12.3$ km) with a 68\% confidence interval of $10.7\pm 0.5$~km ($11.5^{+0.3}_{-0.4}$ km) for $n_{\rm tr,2}$. 

For the phenomenological interaction, the mass-radius relation is much narrower 
than for the chiral interactions because the EOS is much stiffer and uncertainties associated 
with the interaction are unknown. For a typical NS, we find 
a radius range of $11.4-14.3$ km with a 68\% confidence interval of 
$12.7^{+0.7}_{-0.6}$ for $n_{\rm tr,1}$ and a very narrow range of $12.8-12.9$ km 
for $n_{\rm tr,2}$.

In all histograms, we compare our findings to the corresponding envelopes  
of~\cite{Hebeler:2013} for a polytropic expansion with three polytropes  and find 
very good agreement for all interaction models. This suggests that our extension 
is general enough to capture similar effects as the polytropic extension. Our results 
are also consistent with other radius constraints using EOSs obtained with the
AFDMC method~\citep{Steiner:2011ft, Steiner:2014pda}.

\begin{table}
\caption{\label{tab:MmaxNcent}
Maximal NS mass and maximal central densities for all interactions and both transition densities.}
\centering
\begin{tabular}{cc|cc}
\hline
\hline
$n_{\rm tr} \, [\fmiq]$ & Interaction & $M_{\rm max}\, [M_{\odot}]$ & $n_{c, \rm max} \, [n_0]$ \\
\hline
 & TPE-only & 2.01 - 3.63 & 2.8 - 8.6 \\
0.18 & TPE+$V_{\rm E, \mathbbm{1}} $ & 2.01 - 3.66 & 2.7 - 8.9 \\
& AV8'+UIX & 2.01 - 3.57 & 2.9 - 8.4 \\
\hline
& TPE-only & 2.01 - 2.79 & 4.6 - 8.7\\
0.32 & TPE+$V_{\rm E, \mathbbm{1}} $ & 2.01 - 2.81 & 4.5 - 9.3\\
& AV8'+UIX & 2.01 - 2.84  &  4.4 - 8.9\\
\hline
\hline
\end{tabular}
\end{table}

In Table~\ref{tab:MmaxNcent}, we show the maximum masses and the maximal 
central densities for all interactions and both transition densities. The upper limit 
for the maximum mass strongly depends on the chosen transition density but not 
on the interaction. For the smaller transition density, the highest achievable 
maximum mass is $\approx 3.6 $ M$_\odot$, independent of the interaction. For 
the higher transition density, the highest achievable maximum mass is $\approx 2.8 
$ M$_\odot$ for all interactions. The maximal central density reached inside the 
NS, $n_{c,\text{max}}$, that is, the central density in the maximum-mass 
NS, ranges between $2.7-8.9 n_0$ for the lower transition density, and
$4.4-9.3 n_0$ for the larger transition density.

\cite{Kurkela:2014vha}, \cite{Fraga:2015xha}, and \cite{Annala:2017llu} discussed 
the possibility of additionally constraining the EOS with information from pQCD at 
very high densities. Their calculation constrains the pressure at $\epsilon \approx
15000$ MeV/fm$^3$ to lie in the range of $p\approx 2600-4000$~MeV/fm$^3$. 
However, we find that the highest pressure inside any stable NS 
$p_{\rm max} \approx 1570$ MeV/fm$^3$ is significantly smaller. Since we cannot 
infer the EOS at densities above the maximum central density from NS observations, 
it is always possible to find curves for the speed of sound with $0\le c_S^2 \le 1$
that connect our models at the maximal central 
density with the pQCD constraint. 

\subsection{Parameter Correlations}\label{sec:correlations}

\begin{figure*}[t]
\begin{center}
\includegraphics[trim= 0 0 0 0, clip=,height=0.3\textwidth]{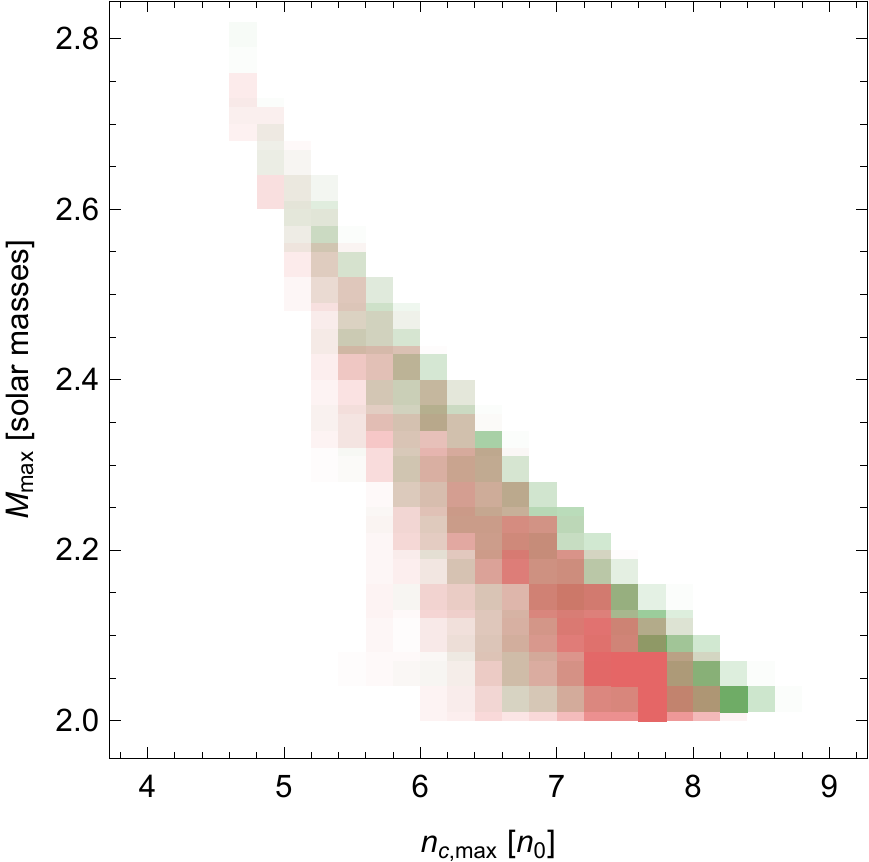} \hspace*{0.2cm}
\includegraphics[trim= 0 0 0 0, clip=,height=0.3\textwidth]{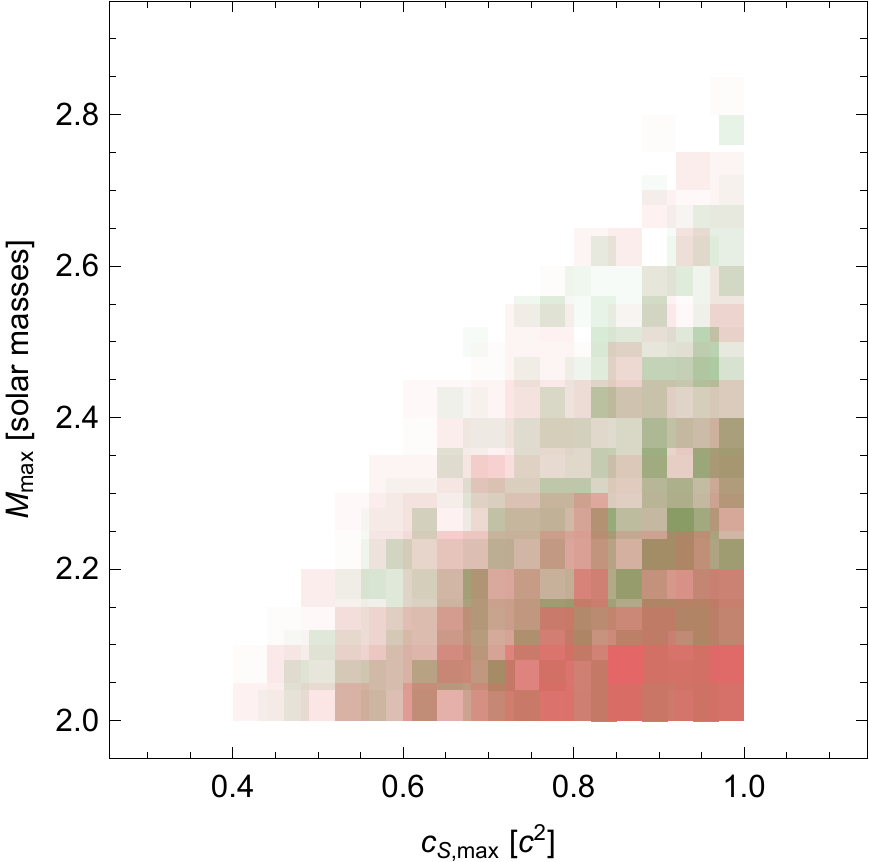}
\end{center}
\caption{\label{fig:parameterCorr}
Histograms for the correlations between $n_{c,\rm max}$ and $M_{\rm max}$ and
$c_{S,\rm max}^2$ and $M_{\rm max}$ for the chiral interactions and $n_{\rm tr,2}$.}
\end{figure*}

In the following, we want to discuss correlations between
$n_{c,\rm max}$ and $M_{\rm max}$, and
$c_{S,\rm max}^2$ and $M_{\rm max}$.
We show histograms for these correlations in Figure~\ref{fig:parameterCorr} for 
the chiral interactions and $n_{\rm tr,2}$, but our observations do not 
change significantly for the other interactions or transition densities.

The correlation between $n_{c,\rm max}$ and $M_{\rm max}$ indicates, as 
expected, that the central densities of the heaviest NSs decrease with 
maximum mass. The same is true for the uncertainties of the central density, 
which also decrease with maximum mass. Furthermore, these central densities
are well below the upper limit established by \cite{Lattimer:2004sa}.

For the correlation between $M_{\rm max}$ and $c_{S,\rm max}^2$ we find that 
the highest masses can only be reached for the stiffest EOS, i.e., EOS with highest
possible $c_{S,\rm max}^2=1$. Furthermore, for every maximum mass there is a 
minimum for $c_{S,\rm max}^2$ that is needed to support this mass. It follows that a 
maximum mass observation will give a lower bound to the maximum speed of 
sound. For the current maximum mass constraint, $M_{\rm max}=2 $ M$_\odot$,
$c_{S,\rm max}^2 \ge 0.4$ for $n_{\rm tr,2}$, which is in excellent agreement with the findings of~\cite{Alsing:2017bbc}. An observation of a $2.4 $ M$_\odot$ NS would require $c_{S,\rm max}^2 \ge 0.6$. 

\section{Smallest possible NS radius}\label{sec:minradius}

In this section we investigate the smallest possible NS radius 
consistent with ab initio calculations of neutron matter and the observation 
of $2 $ M$_\odot$ NSs. This radius is found for the softest possible 
low-density EOS combined with the stiffest possible high-density EOS 
consistent with these constraints~\citep{Koranda:1996jm}.

We again assume that ab initio neutron-matter calculations are valid up to a 
transition density $n_1$, which is at least nuclear saturation density. We choose 
the softest PNM EOS up to $n_1=n_0$ to construct the lower-density part of the 
softest NS EOS. This EOS is given by the lower bound of the uncertainty 
band of the chiral N$^2$LO TPE+$V_{E,\tau}$ interaction (blue band in 
Figure~\ref{fig:nmeos}). Although we did not use this interaction previously, as it 
leads to attractive 3N contributions and negative pressure at $1.5 n_0$, it is the 
most conservative choice to estimate the smallest possible NS radius. 
We then extend this EOS to densities above $n_0$ in the softest way possible 
by setting $c_S=0$ up to a second transition density $n_2$. For the EOS to be 
able to fulfill the second constraint, namely to reproduce a NS of a certain
mass, the high-density part of the EOS has to be as stiff as possible, and above 
$n_2$, we set $c_S=1$. This parameterization is similar to the one explored in
\cite{Alford:2015dpa} and \cite{Alford:2015gna} for $n_{\rm trans}=n_0$, and
leads to the smallest radii consistent with the two constraints, because the radius is 
set by the low-density part of the EOS while the maximum mass is set by the 
high-density part. A softer high-density EOS would require a stiffer low-density 
EOS, which in turn would result in larger radii. 

\begin{figure}[t]
\centering
\includegraphics[trim= 0 0 0 0, clip=,width=0.49\columnwidth]{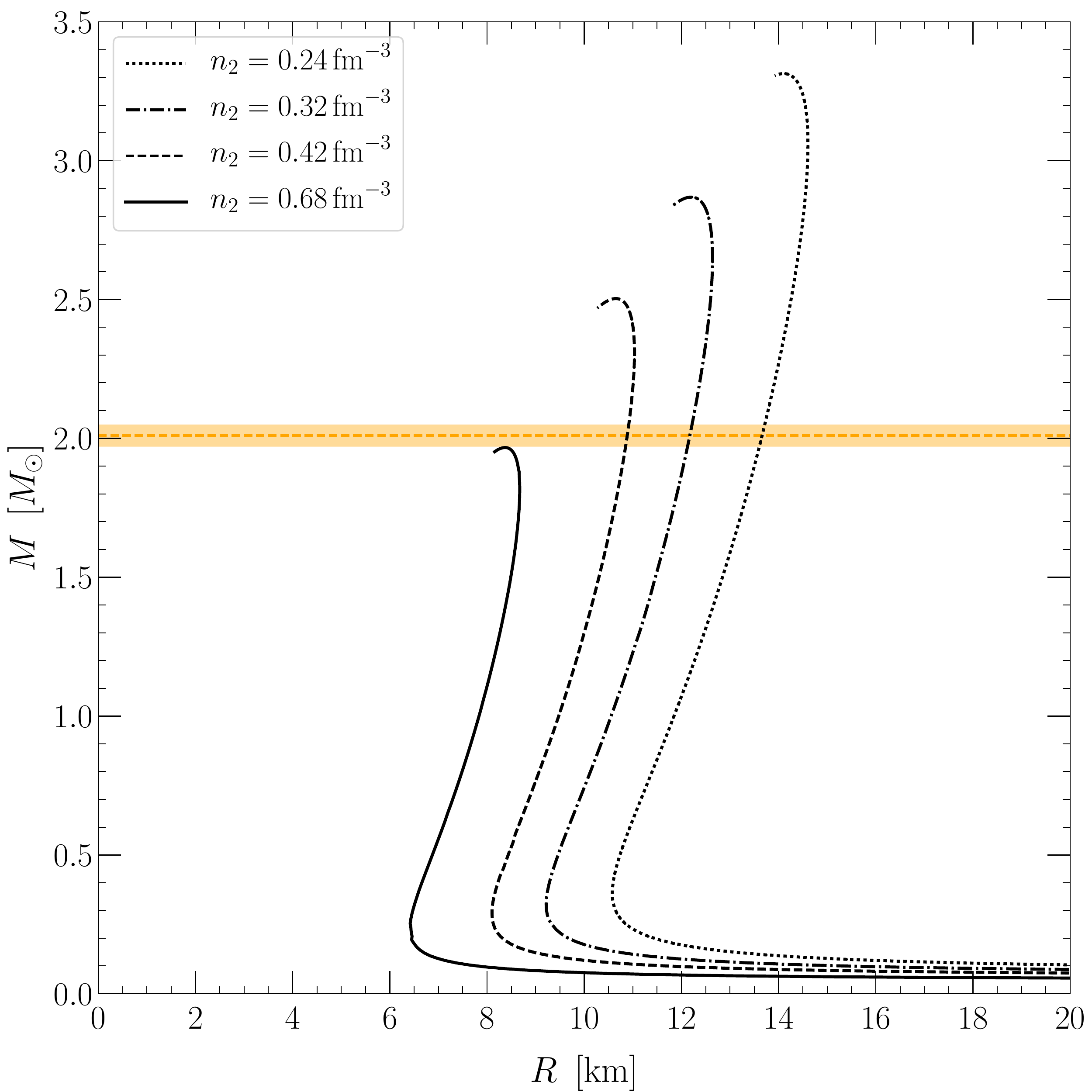} 
\caption{Mass-radius curves for the softest EOS consistent 
with neutron-matter calculations up to saturation density for different densities 
$n_2$. The softest EOS consistent with the observation of a $2 $ M$_\odot$ 
NS is obtained for $n_2=0.68 \fmiq$ (solid black line).
\label{fig:MinR}}
\end{figure}

Changing the transition densities naturally affects the radius: increasing $n_1$
or lowering $n_2$ would increase the radius. The density $n_2$ also determines 
the maximum mass $M_{\rm max}$ that can be supported by this softest EOS. 
Increasing $n_2$ leads to decreasing maximum masses, see Figure~\ref{fig:MinR}. 
We require $M_{\rm max}$ to be at least consistent with the lower uncertainty bound 
of the heaviest observed NS, which is the case for $n_2 = 0.68 \fmiq
=4.25 n_0$ (solid black line). The corresponding curve represents the lowest 
possible NS radii consistent with ab initio neutron-matter calculations 
and current observational constraints on NS masses. This curve implies 
that a typical $1.4$ M$_\odot$ NS has to have a radius larger than 
$8.4$ km. 

If heavier NSs would be observed, $n_2$ has to decrease, which in turn
leads to increasing radii; see Figure~\ref{fig:MinR}. For different $n_2$, the EOS 
constructed here will produce the smallest possible radius that is consistent with 
the corresponding $M_{\rm max}$. If, e.g., a $2.5 $ M$_\odot$ NS was
observed, the radius of a typical NS would have to be larger than 
$10.2$~km.

\section{Impact of additional observations}\label{sec:obs}

\begin{figure*}[t]
\centering
\includegraphics[trim= 0 0 0 0, clip=,width=0.9\textwidth]{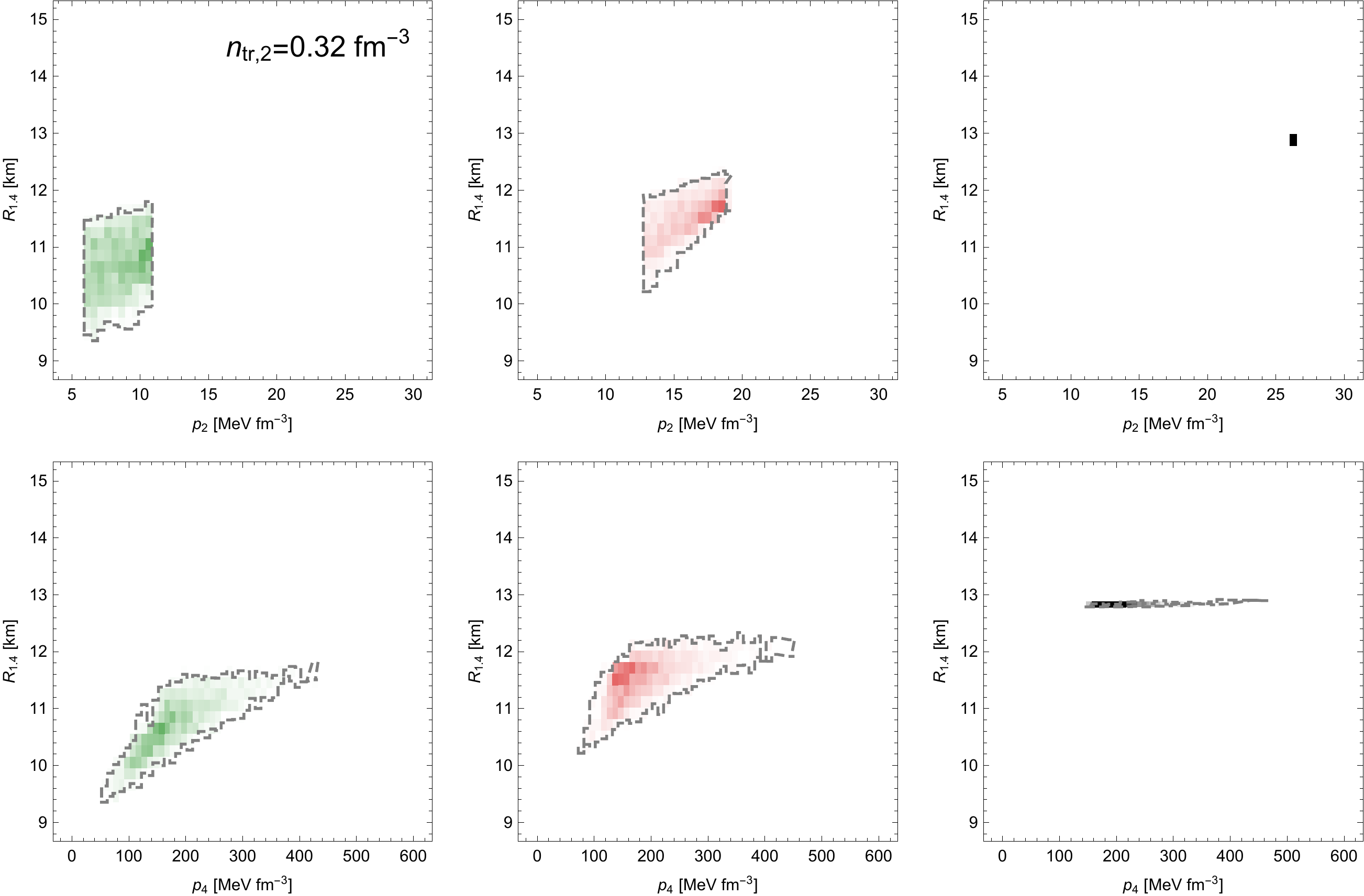} 
\caption{Histograms for the radius of a $1.4$ M$_\odot$ NS, $R_{1.4}$
vs. the pressure at twice saturation density, $p(2n_0)=p_2$ (upper panels), or 
the pressure at four times saturation density, $p(4n_0)=p_4$ (lower panels), for 
all interactions (N$^2$LO TPE-only left, N$^2$LO TPE+$V_{E,\mathbbm{1}}$ 
middle, AV8'+UIX right) and $n_{\rm tr,2}$. 
\label{fig:R1p24plot}}
\end{figure*}

In the following section, we investigate to what extent additional observations 
of NS properties, e.g., NS radii, may help to constrain the 
properties of nuclear interactions. We will assume in this section that we can trust nuclear interactions up to $n_{\rm tr,2}=2n_0$.

For a study of the impact of the recent NS merger observation by the Advanced LIGO collaboration~\citep{LV:2017} using the speed-of-sound extension presented in this paper, please see \cite{Tews:2018iwm}. For studies using different models for the high-density EOS constrained by chiral interactions at low densities, please see \cite{Annala:2017llu}, \cite{Most:2018hfd}, and \cite{Lim:2018bkq}.

\subsection{Observation of compactness}

We begin by assuming that the compactness of a NS has been observed.
It was claimed by~\cite{Hambaryan:2017wvm} that the compactness of the NS RX J0720.4-3125 can be inferred to be $(M/M_{\odot})/(R/{\rm km})=0.105\pm 
0.002$. \cite{Margueron:2017lup} used this information to constrain the 
corresponding NS mass to be $1.33\pm 0.04 $ M$_\odot$ with a radius 
of $12.7\pm 0.3$ km. 

Investigating the effect of such a compactness observation, we find a rather 
broad range for possible NS masses and radii, ranging from 
$M=1.00-1.38 $ M$_\odot$ and $R=9.8-12.8$ km. The chiral 
interactions favor smaller NSs, with the weight of the distribution around 
the point ($11$ km, $1.1 $ M$_\odot$) for the TPE-only and 
($12$ km, $1.2 $ M$_\odot$) for the TPE+$V_{E,\mathbbm{1}}$ interaction. The 
AV8'+UIX interaction leads to a prediction of $M=1.32-1.38 $ M$_\odot$ with a 
radius of $12.8$ km, in good agreement with the prediction of 
\cite{Margueron:2017lup}. 

The observation of the compactness alone is naturally not sufficient to further 
constrain the EOS. However, by observing the compactness of a star with 
known mass, which is equivalent to a radius measurement, additional constraints 
can be found. We discuss this in the next section.

\subsection{Observation of the NS radius}\label{sec:radius}

In the following, we investigate how possible radius measurements by the 
NICER mission will impact our findings for the EOS and the 
mass-radius curve and help to constrain the microscopic equation of state. 
The NICER mission, which was launched in July 2017, is expected to 
measure the compactness (and, thus, radius) of at least three NSs 
with known masses with an accuracy of $5-10 \%$. Among these 
are~\citep{NICER} the NSs 
PSR J1023+0038 with $M=1.71 \pm 0.16$~M$_\odot$~\citep{Deller:2012hp}, 
PSR J0437-4715 with $M=1.44\pm 0.07$~M$_\odot$~\citep{Reardon:2015kba},
and a proposal for using NICER to measure PSR J1614-2230 with a 
mass of $\approx 2.0 $ M$_\odot$~\citep{Miller:2016kae}.

We show histograms for the radius of a typical $1.4$ M$_\odot$ NS, 
$R_{1.4}$, vs. the pressure at two times the nuclear saturation density, $p(2n_0)=p_2$, 
and the pressure at four times the nuclear saturation density, $p(4n_0)=p_4$ in 
Figure~\ref{fig:R1p24plot}, for all interactions and $n_{\rm tr,2}$. All interactions 
predict different $p_2$, as expected, but overlapping distributions for $p_4$.
The chiral interactions predict similar radii around $10-12$ km, and the AV8'+UIX 
interaction predicts a higher radius of $R_{1.4}\approx 12.8$~km, which is not 
consistent with the chiral models; see also Figure~\ref{fig:2DHistogramAV8UIX}. 

\begin{figure*}[t]
\begin{center}
\includegraphics[trim= 0 0 0 0, clip=,width=0.285\textwidth]{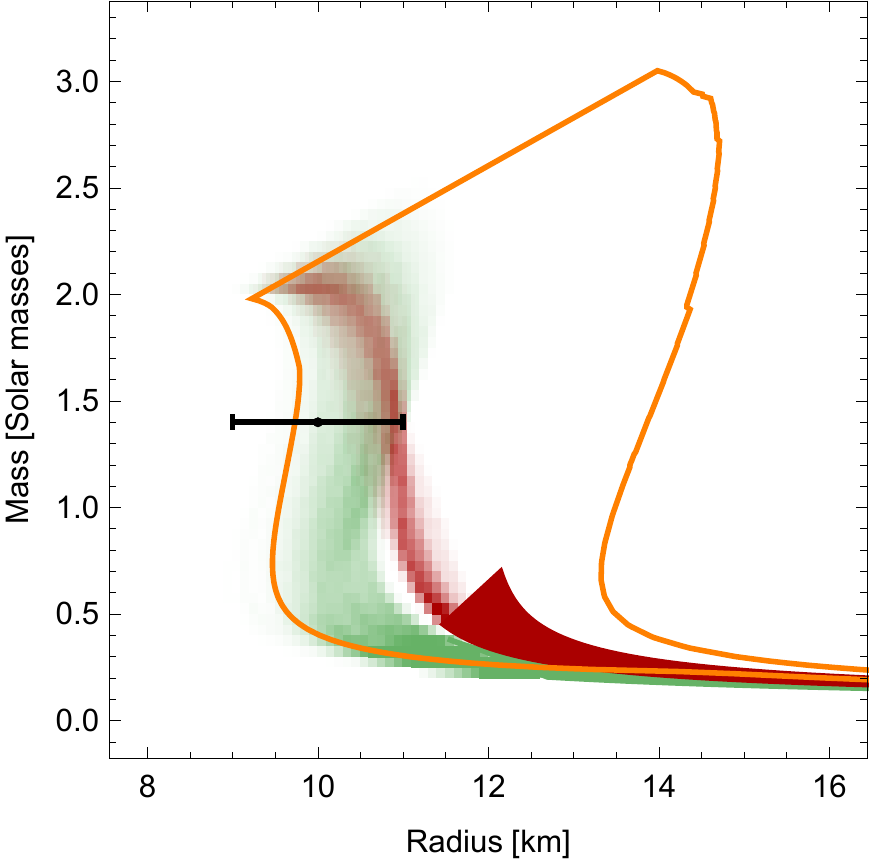} \hspace*{0.2cm}
\includegraphics[trim= 0 0 0 0, clip=,width=0.285\textwidth]{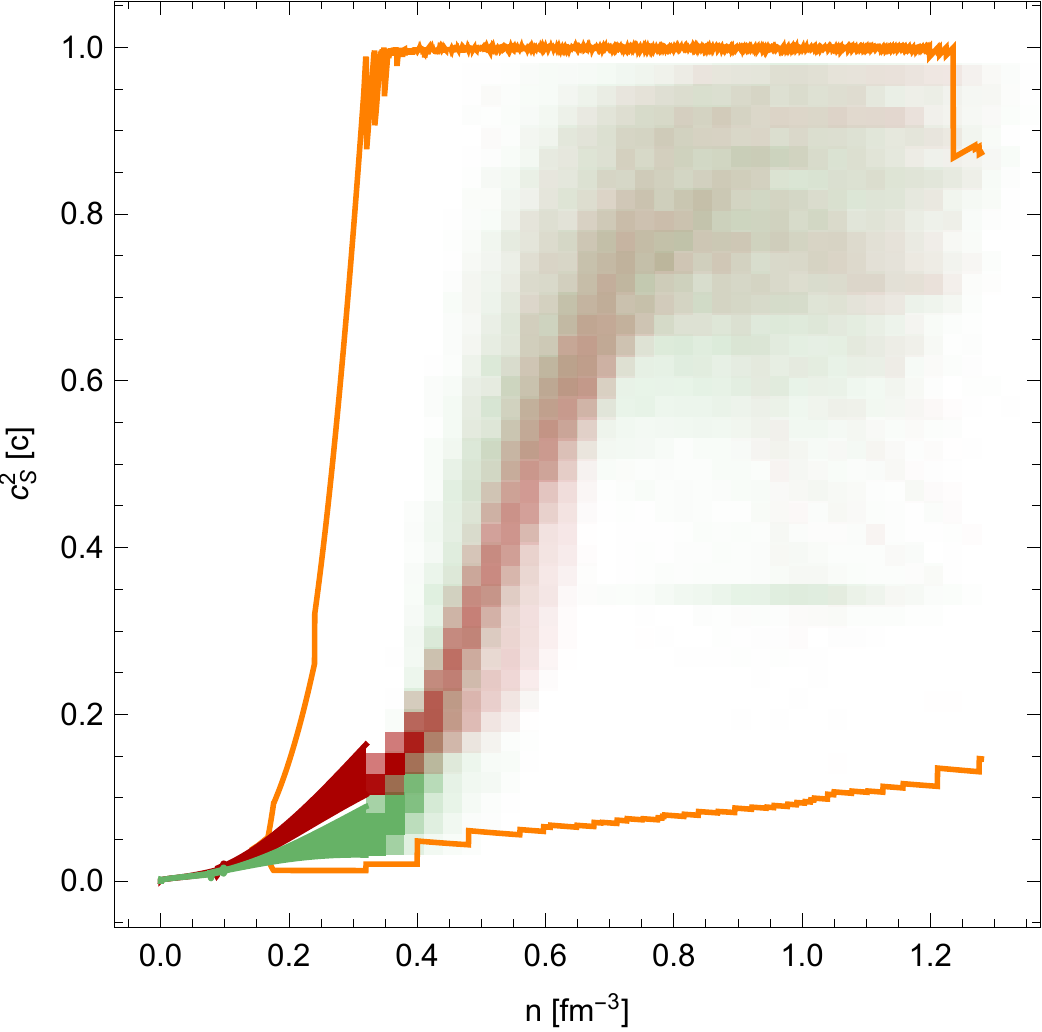} \\
\includegraphics[trim= 0 0 0 0, clip=,width=0.285\textwidth]{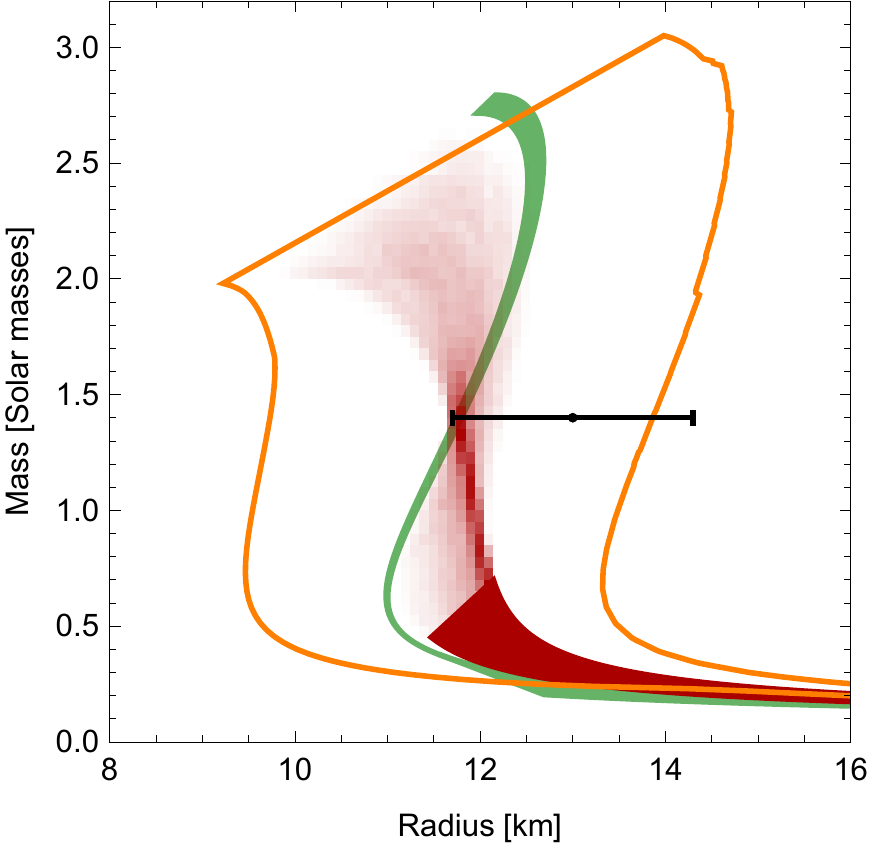} \hspace*{0.2cm}
\includegraphics[trim= 0 0 0 0, clip=,width=0.285\textwidth]{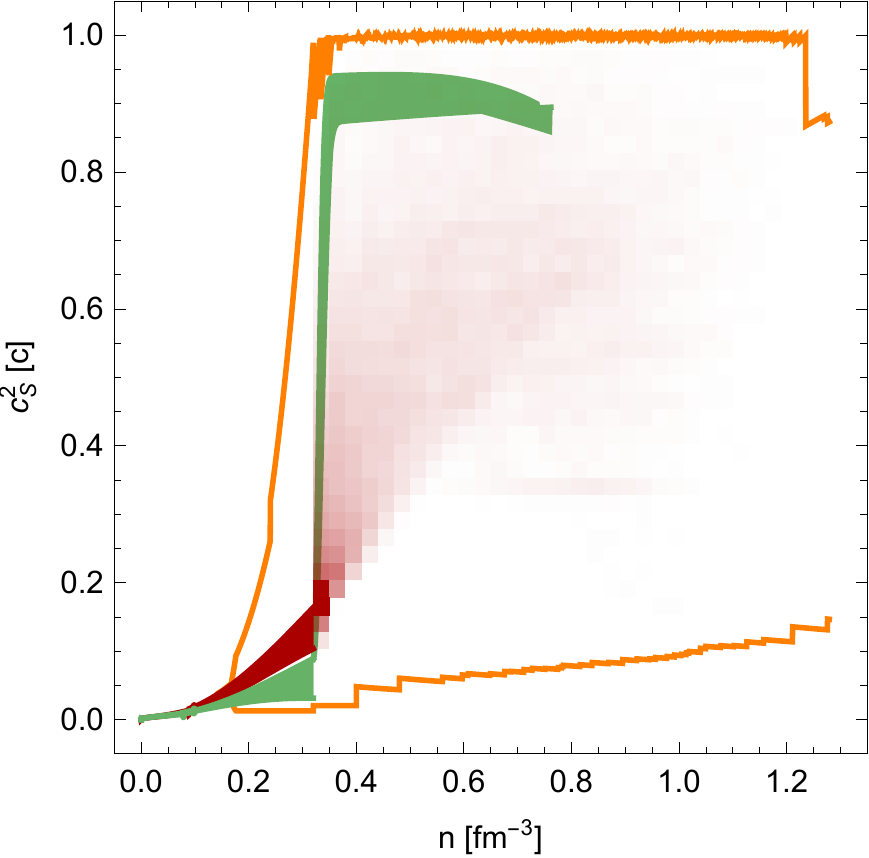} 
\end{center}
\caption{Upper two panels: histograms for the mass-radius relation and the 
speed of sound for the chiral interactions and assuming an observation of a 
NS with $M=1.4$ M$_\odot$ and $R=10\pm 1.0$ km (black point). 
Lower two panels: the same assuming an observation of a NS with 
$M=1.4$ M$_\odot$ and $R=13\pm 1.3$ km.
\label{fig:MRmeasurement}}
\end{figure*}

We now assume that the radius of such a $1.4$ M$_\odot$ NS is 
observed with an accuracy of $10 \%$. We will focus on two extreme cases:
$R_{\rm obs}=(10\pm1.0)$ km, which is the lower bound on the suggested 
radius range of~\cite{Ozel:2016oaf}, and $R_{\rm obs}=(13\pm1.3)$ km, 
which is on the upper end of the currently accepted radius range.
We show the corresponding histograms for the mass-radius relation and the 
speed of sound in Figure~\ref{fig:MRmeasurement}.

The observation of a small $10$ km NS would eliminate a sizable part of the 
stiffer parameterizations with higher $p_2$ and $p_4$ (see also 
Figure~\ref{fig:R1p24plot}) and would allow us to obtain additional constraints on 
the microscopic EOS at $n_{\rm tr,2}$: It would (i) exclude the stiffest chiral 
interactions; (ii) rule out the AV8'+UIX interaction; (iii) reduce the allowed
maximum mass to $\approx 2.5 $ M$_\odot$; and (iv) suggest that the speed of 
sound changes less drastically above $2n_0$ and peaks at densities of 
$\sim 4-5n_0$ with $c_S^2\approx 0.8 \pm 0.1 c^2$. 

The observation of a large $13$ km NS, instead, would exclude the softest 
interactions. For the TPE-only interaction, for example, only a very small fraction 
of parameterizations would survive, and a major fraction of the uncertainty band 
for that interaction could be ruled out. Also, in this case, the speed of sound 
has to increase quickly above $2n_0$.
 
Clearly, observations of more extreme radii and/or with smaller uncertainties 
would present even stronger constraints. For example, an observation of a 
$13$ km NS with 5\% uncertainty would rule out the TPE-only interaction. 
The prospect that NICER may achieve this precision soon with better understood 
systematic errors is exciting for nuclear physics and we eagerly await its results. 

\subsection{Observation of two NS radii}\label{sec:tworadius}
 
 \begin{table}
\caption{\label{tab:p2radius}
Pressure at two times saturation density, $p_2$, for all interactions
and $n_{\rm tr, 2}=0.32 \fmiq$, for hypothetical simultaneous measurements of 
the radii of a $1.4$ M$_\odot$ and a $2.0$ M$_\odot$ NS, $R_{1.4}$ and $R_{2.0}$, respectively.}
\centering
\begin{tabular}{c|cc|cccc}
\hline
\hline
\multicolumn{6}{c}{$p_2\,[\mev \fmiq]$}
\\
\hline
\multicolumn{2}{c}{} & &  \multicolumn{4}{c}{$R_{1.4}\, [\rm{km}]$} \\
\multicolumn{2}{c}{} & & $10\pm 5\%$ & $11\pm 5\%$ & $12\pm 5\%$ & $13\pm 5\%$\\
\hline
\multirow{15}{*}{$R_{2.0} \, [\rm{km}]$} & & TPE-only & 6.5-10.7 & & & \\
& $9\pm 5\%$ & TPE+$V_{\rm E, \mathbbm{1}} $ & & & &\\
& & AV8'+UIX & &  & &\\
\hhline{~------}
& & TPE-only & 5.9 - 10.9 & 5.9 - 10.9 & &\\
& $10\pm 5\%$ & TPE+$V_{\rm E, \mathbbm{1}} $ & 12.8-13.6 & 12.8-18.1 & 14.7-18.9  &\\
& & AV8'+UIX & & & &  \\
\hhline{~------}
& & TPE-only & 5.9 - 10.4 & 5.9 - 10.9  & 9.8-10.8 &\\
& $11\pm 5\%$ & TPE+$V_{\rm E, \mathbbm{1}} $ & & 12.8-18.4 & 12.8-18.9 &\\
& & AV8'+UIX & & & & 26.0\\
\hhline{~------}
& & TPE-only & & 5.9 - 10.9 & 6.1-10.9 &\\
& $12\pm 5\%$ & TPE+$V_{\rm E, \mathbbm{1}} $ & & 12.8-16.8 & 12.8-18.8 &\\
& & AV8'+UIX & & & & 26.0\\
\hhline{~------}
& & TPE-only & & &10.8-10.9 &\\
& $13\pm 5\%$ & TPE+$V_{\rm E, \mathbbm{1}} $ & & & 13.1-18.8 &\\
& & AV8'+UIX & & & & 26.0\\
\hline
\hline
\end{tabular}
\end{table}

  \begin{table}
\caption{\label{tab:p4radius}
Pressure at four times saturation density, $p_4$, for all interactions
and $n_{\rm tr, 2}=0.32 \fmiq$, for hypothetical simultaneous measurements of 
the radii of a $1.4$ M$_\odot$ and a $2.0$ M$_\odot$ NS, $R_{1.4}$ and $R_{2.0}$, respectively.}
\centering
\begin{tabular}{c|cc|cccc}
\hline
\hline
\multicolumn{6}{c}{$p_4\,[\mev \fmiq]$}
\\
\hline
\multicolumn{2}{c}{} & &  \multicolumn{4}{c}{$R_{1.4}\, [\rm{km}]$} \\
\multicolumn{2}{c}{} & & $10\pm 5\%$ & $11\pm 5\%$ & $12\pm 5\%$ & $13\pm 5\%$\\
\hline
\multirow{15}{*}{$R_{2.0} \, [\rm{km}]$} & & TPE-only & 52-92 & & & \\
& $9\pm 5\%$ & TPE+$V_{\rm E, \mathbbm{1}} $ & & & &\\
& & AV8'+UIX & &  & &\\
\hhline{~------}
& & TPE-only & 69-182 & 113-159 & &\\
& $10\pm 5\%$ & TPE+$V_{\rm E, \mathbbm{1}} $ & 72-112  & 84-152 & 107-141 & \\
& & AV8'+UIX & & & &  \\
\hhline{~------}
& & TPE-only & 151-242 & 145-310 & 168-192 & \\
& $11\pm 5\%$ & TPE+$V_{\rm E, \mathbbm{1}} $ & & 122-265  & 117-252 &\\
& & AV8'+UIX & & &  & 147-152\\
\hhline{~------}
& & TPE-only & & 164-428 & 178-428 &\\
& $12\pm 5\%$ & TPE+$V_{\rm E, \mathbbm{1}} $ & & 182-306 & 163-442 &\\
& & AV8'+UIX & & & & 148-285 \\
\hhline{~------}
& & TPE-only & &  & 425-426 &\\
& $13\pm 5\%$ & TPE+$V_{\rm E, \mathbbm{1}} $ & & & 230-446 & \\
& & AV8'+UIX & & &  & 171-461 \\
\hline
\hline
\end{tabular}
\end{table}

As we have shown in the previous section, the observation of a single NS 
radius might prove useful to constrain nuclear interactions, if the observed 
radius is either on the lower or on the upper side of the currently accepted 
range of $12 \pm 2$ km for typical $1.4$ M$_\odot$ NSs. 
We now investigate to which extent the observation of two NS radii 
with $5\%$ uncertainty for stars with known masses can be used to constrain 
both the EOS and nuclear interactions. 

We assume that radii of a $1.4$ M$_\odot$ NS, $R_{1.4}$, and of a 
$2.0 $ M$_\odot$ NS, $R_{2.0}$, have been observed, and present 
the pressure at twice saturation density, $p_2$, in Table~\ref{tab:p2radius} 
and the pressure at four times saturation density, $p_4$, in 
Table~\ref{tab:p4radius} for all interactions and assuming $n_{\rm tr,2}$.

For $n_{\rm tr,2}$, the pressure $p_2$ is set by the nuclear input EOS 
and independent of the extension. As we have shown in 
Section~\ref{sec:csnonconf}, different interactions are compatible with different 
ranges for $R_{1.4}$. The chiral interactions lead to $R_{1.4}\approx10-12$ km
and the AV8'+UIX interaction is only compatible with $R_{1.4}=13$~km $\pm 5\%$. 
If, for instance, $R_{1.4}=10$~km $\pm 5\%$ was observed, the AV8'+UIX 
interaction would be ruled out. 

An additional observation of $R_{2.0}$ would permit further constraints.
For instance, both chiral interactions permit $R_{1.4}=12$ km, but only the 
TPE+$V_{\rm E, \mathbbm{1}} $ interaction could simultaneously lead to 
$R_{2.0}=10$ km. Also, the second radius observation might prove useful to 
constrain $p_4$ and the high-density equation of state. For a single radius
measurement, the predicted ranges for $p_4$ are very large. For example, for
$R_{1.4}=12$ km the predicted $p_4=107-446 \mev \fmiq$. A second 
radius measurement would allow us to clearly reduce this range in most cases.
If, for example, $R_{2.0}=11$ km, the range for $p_4$ would reduce to  
$p_4=117-252 \mev \fmiq$.

The observation of two NS radii could be very useful to constrain both low- and 
high-density EOS, and will hopefully be made available by the NICER mission. 

\section{Summary}\label{sec:summary}

In this work, we used constraints on the neutron-matter EOS at low densities 
and general considerations for the speed of sound in NSs to investigate
the structure of NSs. 

We found that the conformal limit of $c_S^2\leq 1/3$ is in tension with current
nuclear physics constraints and observations of two-solar-mass NSs,
in accordance with the findings of~\cite{Bedaque:2014sqa}.  If the conformal
limit was found to hold at all densities, this would imply that nuclear physics
models break down below $2n_0$.

We then allowed the speed of sound to exceed the conformal limit and used
general considerations about its high-density limit to parameterize the speed of
sound. By using randomly sampled parameter sets and requiring the EOS to 
reproduce two-solar-mass NSs, we computed histograms for the speed 
of sound, the mass-radius relation, and the EOS for microscopic interactions 
from chiral EFT and the AV8'+UIX interaction. We found that the speed of sound 
likely exhibits a sharp increase around $2n_0$ for all interactions under 
consideration. We found that the upper limit on the maximum mass of NSs is $2.9-3.5 $ M$_\odot$, and that radii for typical $1.4$ M$_\odot$ NSs range between 10 and 14 km, in agreement with the results 
of~\cite{Hebeler:2013} and \cite{Steiner:2017vmg}. 

We then studied the minimal possible NS radius consistent with microscopic
ab initio neutron-matter calculations and NS observations, and found 
that a typical $1.4$ M$_\odot$ NS has to have a radius larger than 
$8.4$ km. 

Finally, we studied the impact of additional observations on our models. If the 
compactness of a NS is observed, as suggested 
by~\cite{Hambaryan:2017wvm}, microscopic calculations allow a broad range of 
radii and masses for the corresponding NS. An additional mass 
measurement, i.e, mass and radius are known simultaneously, instead, might put 
tight constraints on the EOS. If the observed radius would be at the 
limits of the currently accepted range of $12 \pm 2$ km, constraints on the 
microscopic interactions would be possible. We have shown that the observation
of two NS radii for NSs with different masses will very likely 
permit tight constraints on nuclear interaction models and the EOS up to several
times nuclear saturation density. 

With the prospect of radius observations becoming available, either from the 
NICER mission or gravitational wave observations from NS mergers 
by the Advanced LIGO collaboration, an exciting era of nuclear astrophysics 
begins. These observations will allow us to finally pin down the EOS of NSs within the coming years.

\begin{acknowledgments}
This work was supported in part by
the National Science Foundation Grant No.~PHY-1430152 
(JINA Center for the Evolution of the Elements),
the U.S.~DOE under grants Nos.~DE-FG02-00ER41132 
and DE-AC52-06NA25396, 
by the NUCLEI SciDAC program, and by the LDRD program at LANL.
Computational resources have also been provided by the J\"ulich
Supercomputing Center.
\end{acknowledgments}

\end{document}